\documentclass[12pt]{article}
\usepackage{amsmath}
\usepackage{amsthm}
\usepackage{amssymb}
\usepackage{graphicx}
\usepackage{enumerate}
\usepackage{natbib}
\usepackage{url} 
\usepackage{booktabs}
\usepackage[toc]{appendix}

\usepackage{color}
\let\oldproofname=\proofname
\renewcommand{\proofname}{\rm\bf{\oldproofname}}

\usepackage{xr}
\newcommand{\blind}{1}

\def \bx {{\boldsymbol x}}

\def \bbeta {\boldsymbol\beta}

\def \bpsi {\boldsymbol\psi}

\def\cero{\mathbf 0}

\def\var{\mbox{Var}}

\begin{document}

\def\spacingset#1{\renewcommand{\baselinestretch}%
{#1}\small\normalsize} \spacingset{1}


\if1\blind
{
  \title{\bf Unifying small area estimators based on area-level and unit-level models through calibration}
  \author{William Acero$^*$\\
    Department of Statistics and Operations Research\\ Complutense University of Madrid
    \vspace{.2cm}\\
    Isabel Molina\thanks{
    This work was supported by the Ministry of Science and Innovation of Spain PID2020-115598RB-I0.} \\
    Institute of Interdisciplinary Mathematics (IMI)\\Department of Statistics and Operations Research\\
    Complutense University of Madrid
    and \vspace{.2cm}\\
    J. Miguel Marín$^*$\\
    Department of Statistics\\
    Carlos III University of Madrid}
    \maketitle
} \fi

\if0\blind
{
  \bigskip
  \bigskip
  \bigskip
  \begin{center}
    {\LARGE\bf Unifying small area estimators based on area-level and unit-level models through calibration}
\end{center}
  \medskip
} \fi

\bigskip
\begin{abstract}
When estimating area means, direct estimators based on area-specific data, are usually consistent under the sampling design without model assumptions. However, they are inefficient if the area sample size is small. In small area estimation, model assumptions linking the areas are used to ``borrow strength'' from other areas. The basic area-level model provides design-consistent estimators but error variances are assumed to be known. In practice, they are estimated with the (scarce) area-specific data. These estimators are inefficient, and their error is not accounted for in the associated mean squared error estimators. Unit-level models do not require to know the error variances but do not account for the survey design. Here we describe a unified estimator of an area mean that may be obtained both from an area-level model or a unit-level model and based on consistent estimators of the model error variances as the number of areas increases. We propose bootstrap mean squared error estimators that account for the uncertainty due to the estimation of the error variances. We show a better performance of the new small area estimators and our bootstrap estimators of the mean squared error. We apply the results to education data from Colombia.
\end{abstract}

\noindent%
\vfill

\newpage
\spacingset{1} 

\section{Introduction}
\label{sec:intro}

When estimating linear characteristics of a finite population, survey practitioners rely on expansion estimators, obtained as weighted sums of the sample observations using the survey weights, which are often previously calibrated.
Design-based inference provides estimators with good properties under the sampling replication mechanism. A good property of common design-based estimators, such as the usual expansion estimators, is design consistency as the sample size increases, which holds without any model assumption for the values of the target variable.
Since the survey sample size is typically planned to achieve a desired level of precision when estimating at the population level or for certain subpopulations, design-based estimators perform well for those planned domains.


Often, after the survey is conducted, disaggregated estimates are required for other subpopulations or areas, for which the survey was not originally planned. In this case, the expansion estimators obtained with the area-specific survey data, called direct estimates, may lack precision due to small area sample sizes. Small area estimation (SAE) methods address the problem of data scarcity by making model assumptions about the study variable. These assumptions often link all the target areas through common parameters. These common parameters are estimated with the survey data from all the areas, leading to ``indirect'' estimators of the target area characteristic that ``borrow strength'' from related areas.
Model-based estimators typically improve the precision of direct estimators, provided that model assumptions hold.

When estimating small area means, two well-known models are often used. The first one is the area-level model proposed by \cite{fay1979estimates}, referred to as the FH model, which
uses only aggregated data at the area level.
Specifically, this model linearly relates the direct estimators of the target area characteristics to a vector of true area-level means of auxiliary variables, and includes area effects and model errors. Based on this model, empirical best linear unbiased predictors (EBLUPs) are obtained.
Direct estimates are previously calculated using the unit-level data on the target variable from a survey, whereas the true area means of the auxiliary variables are typically obtained from census data.

The second model is the unit-level model proposed by \cite{battese1988error}, referred to as the BHF model, which assumes a linear regression model for the individual values of the target variable in terms of the available unit-level covariates and includes random area effects. Linear characteristics are then estimated with the EBLUPs obtained under this model, which  also requires the population means of the covariates for the area, in addition to the unit-level survey data on the target variable and the covariates. Both FH and BHF models belong to the class of linear mixed models. Their main difference is that FH model is fit using only area-level data, whereas BHF model is fit with the unit-level survey data.

An important advantage of FH model over BHF one is that the resulting EBLUPs based on FH model may use the survey weights through the direct estimates, which act as response variables in the FH model. As consequence, the EBLUPs based on FH model are design consistent as the area sample size becomes large, unlike EBLUPs of area means based on BHF model, which ignore the sampling design.

\cite{you2002pseudo} obtained a Pseudo EBLUP for a small area mean under BHF model that is design consistent as the area sample size increases. They provided an analytical approximation for the mean square error (MSE) of the Pseudo EBLUP, as well as an estimator of the MSE. Moreover, if survey weights for the units in each area are calibrated to add up to the known population count of that area, the Pseudo EBLUPs of the area totals automatically satisfy the convenient self-benchmarking property of adding up to the survey regression estimator of the population total.

A well-known issue of the FH model is that the variances of the model errors are assumed to be known. Since these variances vary across areas, they cannot be estimated from the available area-level data. This problem is typically solved by replacing these variances with some estimates of the sampling variances of the direct estimators, obtained from the unit-level area-specific survey data. These estimates are then treated as the true error variances in the FH model. However, they are also subject to high sampling variability for areas with small sample sizes, and this variability is typically ignored.

A common approach to avoid the problem of the high variability of the (direct) estimators of the sampling variances is to smooth them using a previous model for the estimated variances in terms of the same or different predictors. However, the choice of predictors, which tunes the  smoothing level of the final variance estimates, obtained as predicted values in this preliminary model, is often unclear and arbitrary. Note that the purpose of this preliminary model is not to find the estimated variances that best fit the direct variance estimates, as that would eliminate the smoothing effect.

\cite{prasad1990estimation} obtained a second-order approximation for the MSE of the EBLUP of a small area mean under the FH model, as the number of areas tends to infinity. They also provided an estimator of the MSE with second-order bias, as the number of areas increases. However, this estimator again assumes that the true error variances in the FH model are known. When these true error variances are replaced by the estimated sampling variances of direct estimators, the resulting Prasad-Rao MSE estimator might underestimate the true MSE. Consequently, gains of the usual EBLUPs based on FH model with respect to direct estimators might be overstated in the areas with small sample sizes.

\cite{bell2008examining} and \cite{rivest2003mean} studied the effect of replacing model error variances with estimated sampling variances of direct estimators in the MSE of the EBLUP based on FH model. They considered simple random sampling within the areas and assumed that the other FH model parameters were known. Simulation results of \cite{rivest2003mean} indicate that the Prasad-Rao MSE estimator can underestimate the true MSE for areas with small sample sizes. \cite{rivest2003mean} proposed an
estimator of the MSE that accounts for the uncertainty due to the estimation of these variances. \cite{wang2003mean} provided another estimator of the MSE when all the model parameters are estimated, using certain moment estimators, but further research is needed to extend their results to other sampling designs and fitting methods. \cite{fay2018further} proposed to consider the survey-regression estimator as direct estimator in the FH model, when unit-level data are available, and used a smoothed version of its estimated variance as the true error variance. He found a better performance of the resulting EBLUP for some simulation scenarios in contrast to using those using customary direct estimators but, again, the smoothing procedure is not clear. \cite{Rao2023} studied the relative efficiency of EBLUPs obtained under Fay's proposed model (referred to as \textit{hybrid model}) compared to the EBLUPs based on the BHF unit-level model. They obtained MSE ratios of EBLUPs under BHF model by those under the hybrid model close to one, showing no major gain in efficiency.


In this paper, we demonstrate how to obtain the area-level FH model through aggregation in the unit-level BHF model, utilizing the survey weights and calibration at the area totals of the covariates. The aggregation process provides a correct specification for the error variances in the aggregated FH model in terms of the variance of the unit-level model errors in the BHF model. This implies that the FH model parameters, including the heteroscedastic error variances, can be estimated using the unit-level data on the target variable and the covariates from the survey. Utilizing unit-level data leads to increased precision due to the larger number of observations used to fit the model.

In cases where no access is provided to the unit-level survey data, the FH model with the same specification of error variances can be also fitted with the area-level data. In both cases, we propose bootstrap MSE estimators that account for the uncertainty due to the estimation of error variances, applicable to general fitting methods and sampling designs.

We compare the usual EBLUPs based on the FH model, which use estimated sampling variances as the true error variances, with the EBLUPs obtained by fitting the unit-level BHF model that leads to the FH model through aggregation. Subsequently, we illustrate that the customary EBLUPs based on FH model result in greater MSEs, especially for areas with the smallest sample sizes. Finally, we investigate the underestimation of the MSE when using the Prasad-Rao MSE estimator, replacing true error variances with estimated sampling variances.
We show that our bootstrap MSE estimators provide better approximations to the true MSEs, even for very small area sample sizes.

\section{Area-level FH model}\label{sec2}

Let $U$ be a finite population of size $N$, composed of $D$ non-overlapping areas or domains $U_1,\ldots,U_D$, each with sizes $N_1,\ldots, N_D$. Let $y_{di}$ represent the value of the study variable for the $i$th unit within the $d$th area, $i=1,\ldots, N_d$, and $d=1,\ldots,D$. The objective is to estimate the area
means $\mu_d=N_d^{-1}\sum_{i=1}^{N_d} y_{di}$, for $d=1,\ldots, D$.

To estimate $\mu_d$ for each $d$, we use data from a survey conducted on the entire population, measuring the variable of interest. In this survey, a sample $s_d\subseteq U_d$ of size $n_d>0$ with $n_d\leq N_d$, is assumed to be drawn independently from each area $d$, for $d=1,\ldots,D$, according to a given sampling design. The overall sample is denoted as $s=s_1\cup \cdots \cup s_D$ with a total sample size of $n=\sum_{d=1}^D n_d$. From the $n_d$ survey observations in area $d$, a direct estimator $\hat{\mu}_d^{DIR}$ of $\mu_d$ is obtained for $d=1,\ldots,D$. The problem arises when direct estimators $\hat{\mu}_d^{DIR}$ for some areas exhibit unacceptable sampling errors, due to small $n_d$.

\cite{fay1979estimates} addressed this problem by considering the model known as FH model and defined in two stages. In the first stage, the true area mean $\mu_d$ were assumed to vary linearly with respect to a vector $\bar{\boldsymbol{X}}_d$ with the area means of $p<n$ covariates:
\begin{equation}
	\mu_d = \bar{\boldsymbol{X}}_d' \boldsymbol{\beta} + u_d,\quad u_d \stackrel{iid}{\sim} N(0, \sigma_u^{2})\quad d=1,\ldots, D.
	\label{linkingModel}
\end{equation}
Here, $\boldsymbol{\beta}$ is a $p\times1$ vector of unknown regression coefficients, and $u_d$, $d=1,\ldots,D$, are known as area effects, with $\sigma_u^{2}>0$ being unknown. This model, known as the \textit{linking model}, allows to ``borrow strength'' by linking all the areas through the common $\boldsymbol{\beta}$.

In model \eqref{linkingModel}, the true means $\mu_d$, are not observable but we consider that direct estimators $\hat{\mu}_d^{DIR}$ of $\mu_d$ are available from a survey. These direct estimators have non-negligible sampling variances, which vary by areas because the area sample sizes $n_d$ are generally different.
To account for the sampling error of the direct estimators, at the second stage, \cite{fay1979estimates} assumed the following model, known as the \textit{sampling model},
\begin{equation}
	\hat{\mu}_d^{DIR} = \mu_d + e_d, \quad e_d \stackrel{ind}{\sim} N(0, \psi_d), \quad d=1,\ldots,D.
	\label{samplingModel}
\end{equation}
Here, the $e_d$'s represent the sampling errors, assumed to be independent of the area effects $u_d$, and with error variances $\psi_d>0$ assumed to be known for $d=1,\ldots, D$.
By combining \eqref{linkingModel} and \eqref{samplingModel}, the FH model can be expressed as a single linear mixed model,
\begin{equation}
	\hat{\mu}_d^{DIR} = \bar{\boldsymbol{X}}_d' \boldsymbol{\beta} + u_d + e_d,\quad d=1,\ldots, D.
	\label{mixMod}
\end{equation}

Let $\bpsi=(\psi_1,\ldots,\psi_D)'$ be the vector of error variances. 
The best predictor (BP) of the mixed effect $\mu_d=\bar{\boldsymbol{X}}_d' \boldsymbol{\beta} + u_d$ under the FH model \eqref{mixMod} is the predictor $\tilde\mu_d$ obtained in terms of the available data $\boldsymbol{y}=(\hat{\mu}_1^{DIR},\ldots, \hat{\mu}_D^{DIR})'$ that minimizes the MSE  under the FH model \eqref{mixMod}, $\mbox{MSE}(\tilde\mu_d)=E[(\tilde\mu_d-\mu_d)^2]$. Under normality of $u_d$ and $e_d$, the BP of $\mu_d=\bar{\boldsymbol{X}}_d' \boldsymbol{\beta} + u_d$ is
\begin{equation}
\tilde{\mu}_d(\bbeta,\sigma_u^2,\bpsi) = \gamma_d(\sigma_u^2,\bpsi) \hat{\mu}_d^{DIR} +  (1-\gamma_d(\sigma_u^2,\bpsi)) \bar{\boldsymbol{X}}_d' \boldsymbol{\beta}, \quad \gamma_d(\sigma_u^2,\bpsi)=\frac{\sigma_u^2}{\sigma_u^2+\psi_d}.
\label{bpFayModel2}
\end{equation}
According to \eqref{bpFayModel2}, the BP of $\mu_d$ under the FH model is a weighted average of the direct estimator $\hat{\mu}_d^{DIR}$ and the regression-synthetic estimator $\bar{\boldsymbol{X}}_d' \boldsymbol{\beta}$, tending to $\hat{\mu}_d^{DIR}$ as $\psi_d$ goes to zero (equivalently, as $n_d$ grows) and to the regression-synthetic estimator otherwise.

The MSE of the BP $\tilde{\mu}_d(\bbeta,\sigma_u^2,\bpsi)$ obtained with the true values of $\bbeta$, $\sigma_u^2$ and $\bpsi$ satisfies
\begin{equation*}
\mbox{MSE}[\tilde{\mu}_d(\bbeta,\sigma_u^2,\bpsi)] = \gamma_{d}(\sigma_u^2,\bpsi)\psi_d < \psi_d=\mbox{\var}(\hat{\mu}_d^{DIR}|\mu_d);
\end{equation*}
hence, the BP $\tilde{\mu}_d(\bbeta,\sigma_u^2,\bpsi)$ is more efficient than the direct estimator $\hat{\mu}_d^{DIR}$, for all $d=1,\ldots,D$. However, in practice, $\boldsymbol{\beta}$, $\sigma_{u}^{2}$ and the error variances $\psi_d$, $d=1,\ldots,D$, are unknown.

Assuming that $\sigma_{u}^{2}$ and $\psi_d$, $d=1,\ldots,D$, are known, but avoiding normality in \eqref{linkingModel} and \eqref{samplingModel}, \cite{fay1979estimates} obtained the best linear unbiased predictor (BLUP) of the mixed effect $\mu_d=\bar{\boldsymbol{X}}_d' \boldsymbol{\beta} + u_d$, which is the optimal predictor in terms of minimizing the MSE under the FH model \eqref{mixMod}, among the class of predictors $\tilde\mu_d$ that are linear on $\boldsymbol{y}=(\hat{\mu}_1^{DIR},\ldots, \hat{\mu}_D^{DIR})'$ and unbiased in the sense $E[\tilde\mu_d-\mu_d]$. The BLUP of $\mu_d=\bar{\boldsymbol{X}}_d' \boldsymbol{\beta} + u_d$ equals the BP $\tilde\mu_d(\bbeta,\sigma_u^2,\bpsi)$, when the true $\bbeta$ is replaced by the weighted least squares (WLS) estimator obtained from area-level data $\{(\hat{\mu}_d^{DIR},\bar{\boldsymbol{X}}_d),\ d=1,\ldots,D\}$ and given by
\begin{equation}
\tilde{\boldsymbol{\beta}}_{FH}(\sigma_u^2,\bpsi) = \left( \sum_{d=1}^{D} \gamma_d(\sigma_u^2,\bpsi) \bar{\boldsymbol{X}}_d\bar{\boldsymbol{X}}_d' \right)^{-1} \sum_{d=1}^{D}\gamma_d(\sigma_u^2,\bpsi) \bar{\boldsymbol{X}}_d \hat{\mu}_d^{DIR}.
\label{BetaFH}
\end{equation}
Therefore, for $\sigma_u^2$ and $\bpsi$ known, the BLUP of $\mu_d$ under the FH model \eqref{mixMod} is given by
\begin{equation}
\tilde{\mu}_d^{FH}(\sigma_u^2,\bpsi)  = \tilde{\mu}_d\left(\tilde{\boldsymbol{\beta}}_{FH}(\sigma_u^2,\bpsi),\sigma_u^2,\bpsi\right) =\gamma_d(\sigma_u^2,\bpsi) \hat{\mu}_d^{DIR} +  \left(1-\gamma_d(\sigma_u^2,\bpsi)\right) \bar{\boldsymbol{X}}_d' \tilde{\boldsymbol{\beta}}_{FH}(\sigma_u^2,\bpsi).
\label{blupFayModel}
\end{equation}

For known $\psi_d$, $d=1,\ldots, D$, a consistent estimator of $\sigma_{u}^{2}$ may be obtained by maximum likelihood (ML), restricted ML (REML) or by a moments method, such as the one proposed by \cite{fay1979estimates}, known as FH fitting method. In terms of relative efficiency, these three estimation methods are equivalent \citep[see][sec. 6.1.3]{rao2015small}.

Replacing a consistent estimator $\hat{\sigma}_{u}^{2}(\bpsi)$ of $\sigma_{u}^{2}$ as $D\to\infty$ in $\eqref{blupFayModel}$, we obtain the empirical BLUP (EBLUP) of $\mu_d$. Denoting $\hat{\boldsymbol{\beta}}_{FH}(\bpsi)=\tilde{\boldsymbol{\beta}}_{FH}\left(\hat\sigma_u^2(\bpsi),\bpsi\right)$ and $\hat\gamma_d(\bpsi)=\gamma_d(\hat\sigma_u^2(\bpsi),\bpsi)$, the EBLUP of $\mu_d$ based on the FH model \eqref{mixMod} with error variances $\psi_d$, $d=1,\ldots,D$, reads
\begin{equation}
\hat{\mu}_d^{FH}(\bpsi)= \tilde{\mu}_d^{FH}\left(\hat\sigma_u^2(\bpsi),\bpsi\right)=\hat{\gamma}_d(\bpsi) \hat{\mu}_d^{DIR} +  \left(1-\hat{\gamma}_d(\bpsi)\right) \bar{\boldsymbol{X}}_d' \hat{\boldsymbol{\beta}}_{FH}(\bpsi),\ \hat\gamma_d(\bpsi)=\frac{\hat\sigma_u^2(\bpsi)}{\hat\sigma_u^2(\bpsi)+\psi_d}.
\label{eblupFayModel}
\end{equation}
Since, for $\bpsi$ known, $\hat{\sigma}_{u}^{2}(\bpsi)$ and $\hat\bbeta_{FH}(\bpsi)$ are consistent estimators of $\sigma_{u}^{2}$ and $\bbeta$, respectively, as $D\to\infty$, we expect $\hat{\mu}_d^{FH}(\bpsi)$ to be close to the BP $\tilde{\mu}_d(\bbeta,\sigma_u^2,\bpsi)$ for $D$ sufficiently large, which is more efficient than the direct estimator $\hat{\mu}_d^{DIR}$.

Assuming normality of $u_d$ and $e_d$, known and uniformly bounded error variances $\psi_d$, $d=1,\ldots,D$, and
$\sup_{1\leq d\leq D}h_{dd}=O(D^{-1})$, for $h_{dd}=\bar{\boldsymbol{X}}_d'\left(\sum_{d=1}^D\bar{\boldsymbol{X}}_d\bar{\boldsymbol{X}}_d'\right)^{-1}\bar{\boldsymbol{X}}_d$, $d=1,\ldots,D$, \cite{prasad1990estimation} approximated the MSE of $\hat{\mu}_d^{FH}(\bpsi)$ in \eqref{eblupFayModel} with
\begin{eqnarray}
\mbox{MSE}_{PR}[\hat{\mu}_{d}^{FH}(\bpsi)] &=&g_{1d}(\sigma _{u}^{2},\bpsi)+g_{2d}(\sigma
	_{u}^{2},\bpsi)+g_{3d}(\sigma_{u}^{2},\bpsi),
\label{MSEFH}
\end{eqnarray}
where
\begin{eqnarray*}
	g_{1d}(\sigma _{u}^{2},\bpsi) &=&\gamma_{d}(\sigma_u^2,\bpsi)\psi _{d}, \\
	g_{2d}(\sigma _{u}^{2},\bpsi) &=&\left(1-\gamma_{d}\sigma_u^2,\bpsi)\right)^{2}\bar{\boldsymbol{X}}_d^{\prime }%
	\left(\sum_{d=1}^{D}\frac{\bar{\boldsymbol{X}}_d\bar{\boldsymbol{X}}_d^{\prime }}{%
		\psi_{d}+\sigma _{u}^{2}}\right) ^{-1}\bar{\boldsymbol{X}}_d, \\
	g_{3d}(\sigma_{u}^{2},\bpsi) &=& \left(1-\gamma_d(\sigma_u^2,\bpsi)\right)^2\frac{\overline{\mbox{var}}(\hat{\sigma}_{u}^{2})}{\sigma^{2}_{u} + \psi_d}.
\end{eqnarray*}
Here, $\overline{\mbox{var}}(\hat{\sigma}_{u}^{2})$ denotes the asymptotic variance of $\hat{\sigma}_{u}^{2}(\bpsi)$, which depends on the procedure employed to estimate $\sigma_u^2$. The term $g_{1d}(\sigma _{u}^{2},\bpsi)$ accounts for the uncertainty due to the prediction of the area effect $u_d$ and is $O(1)$ as $D\to \infty$, whereas $g_{2d}(\sigma _{u}^{2},\bpsi)$ and $g_{3d}(\sigma_{u}^{2},\bpsi)$ are the uncertainty due to the estimation of $\bbeta$ and $\sigma_u^2$, respectively, and are both $O(D^{-1})$ as $D\to \infty$. The approximation \eqref{MSEFH} is second-order correct, meaning that
$$
\mbox{MSE}[\hat{\mu}_{d}^{FH}(\bpsi)]=\mbox{MSE}_{PR}[\hat{\mu}_{d}^{FH}(\bpsi)]+o(D^{-1}).
$$

Based on \eqref{MSEFH}, the authors provided a second-order unbiased estimator of the MSE of $\hat{\mu}_{d}^{FH}(\bpsi)$, that is, with $o(D^{-1})$ bias, which again depends on the estimation method used for $\sigma_u^2$. If $\hat{\sigma}_{u}^{2}(\bpsi)$ is the REML estimator of $\sigma_u^2$, the second-order unbiased MSE estimator is
	\begin{equation}
	\mbox{mse}_{PR}[\hat{\mu}_d^{FH}(\bpsi)] =  g_{1d}(\hat{\sigma}_{u}^{2}(\bpsi),\bpsi) + g_{2d}(\hat{\sigma}_{u}^{2}(\bpsi),\bpsi) + 2g_{3d}(\hat{\sigma}_{u}^{2}(\bpsi),\bpsi).
	\label{msePR}
\end{equation}	

In practice, the error variances in $\bpsi=(\psi_1,\ldots,\psi_D)'$ are unknown. Hence, we need to estimate them to obtain the EBLUPs based on FH model, $\hat{\mu}_d^{FH}(\bpsi)$. A common approach is to replace $\psi_d$ with the estimated design-based variance of the direct estimator, $\psi_{d0}=\widehat{\var }_{\pi}(\hat\mu_d^{DIR}|\mu_d)$, which is obtained from the $n_d$ unit-level survey observations from area $d$ and hence it is also ``direct''. Defining $\bpsi_0=(\psi_{10},\ldots,\psi_{D0})'$, the FH estimator obtained by setting $\bpsi=\bpsi_0$ is denoted here as $\hat{\mu}_d^{FHD}=\hat{\mu}_d^{FH}(\bpsi_0)$.

Smoothed versions of $\psi_{d0}$, $d=1,\ldots,D$, obtained as fitted values from a preliminary regression model for them are often employed to avoid the high instability of $\psi_{d0}$. However, no clear procedure for selecting the predictors in that model seems to be given, and this determines the level of smoothing of the variance estimators.

Moreover, using the approximation $\mbox{MSE}_{PR}(\hat{\mu}_{d}^{FHD})$ obtained by setting $\bpsi=\bpsi_0$ in \eqref{MSEFH} is likely to understate the true MSE of $\hat{\mu}_d^{FHD}=\hat{\mu}_{d}^{FH}(\bpsi_0)$ for small $n_d$, a problem that inherits the associated MSE estimator, $\mbox{mse}_{PR}(\hat{\mu}_{d}^{FHD})$, obtained by setting $\bpsi=\bpsi_0$ in \eqref{msePR}. They are not accounting for the uncertainty due to the estimation of $\psi_d$ by $\psi_{d0}$, $d=1,\ldots,D$, which is not negligible for $n_d$ small, as will be illustrated in the simulation experiments of Section \ref{sec5}. Actually, as will be also shown in our simulations, in areas with a small sample size $n_d$, since the true MSE of $\hat{\mu}_d^{FHD}$ is larger than expected, $\hat{\mu}_d^{FHD}$ might be even less efficient than the corresponding direct estimator $\hat{\mu}_d^{DIR}$.

\section{Unit-level BHF model}\label{sec3}

\cite{battese1988error} proposed a nested error linear regression model for data at the unit level, in the context of
estimating mean acreage under a crop for counties (small areas) in Iowa. The model, assumed for each population unit and known as BHF model, is given by
\begin{equation}
	y_{di} = \boldsymbol{x}_{di}'\boldsymbol{\beta}+ u_d + e_{di}, \quad i=1,\ldots,N_d, \ d=1,\ldots,D,
	\label{UnitlinearMixModel}
\end{equation}
where $\boldsymbol{x}_{di}=(x_{di1},\ldots,x_{dip})'$ contains the unit-specific values of $p<n$ auxiliary variables, $\boldsymbol{\beta}$ is a $p\times1$ vector of unknown regression coefficients, $u_{d}$ is the random effect of area $d$, assumed to satisfy $u_{d}\stackrel{iid}{\sim} N(0, \sigma_{u}^{2})$, $d=1,\ldots, D$, and $e_{di}$ are the unit model errors, satisfying $e_{di}\stackrel{iid}{\sim} N(0, \sigma_{e}^{2})$, $i=1,\ldots, N_d$, and all independent of the area effects.
If sample selection bias is absent, the sample measurements $(y_{di}, \boldsymbol{x}_{di})$, $i\in s_d$, $d=1,\ldots,D$, follow the same population model given in \eqref{UnitlinearMixModel}.

\cite{battese1988error} obtained the EBLUP of $\mu_d=\bar{\boldsymbol{X}}_d'\boldsymbol{\beta}+u_d$ under the model \eqref{UnitlinearMixModel} and \cite{prasad1990estimation} provided a second-order approximation for the MSE of the EBLUP, together with a second-order unbiased MSE estimator.
However, the EBLUP of $\mu_d$ under the BHF model \eqref{UnitlinearMixModel} ignores the sampling design. As a consequence, it is not design consistent as the area sample size $n_d$ grows, unlike the EBLUP based on the FH model \eqref{mixMod}.

\cite{prasad1999robust} introduced a Pseudo EBLUP for a small area mean that uses the survey weights and is design-consistent as the area sample size $n_d$ increases.
Let $w_{di}>0$ be the survey weight for unit $i$ in the sample from area $d$ and $w_{d\cdot}=\sum_{i\in s_d}w_{di}$.
The survey-weighted direct estimator $\bar{y}_{dw} =w_{d\cdot}^{-1}\sum_{i\in s_d} w_{di}y_{di}$ is design-consistent as $n_d$ increases.
They noted that taking weighted average in BHF model \eqref{UnitlinearMixModel} over the sample units in area $d$ results in the following aggregated area-level model,
\begin{equation}
	\bar{y}_{dw} = \bar{\boldsymbol{x}}_{dw}'\boldsymbol{\beta}+ u_d + \bar{e}_{dw}, \quad d=1,\ldots,D,
	\label{AreaLevelModel}
\end{equation}
where $\bar{\boldsymbol{x}}_{dw}=w_{d\cdot}^{-1}\sum_{i\in s_d}w_{di}\boldsymbol{x}_{di}$ and $\bar{e}_{dw}=w_{d\cdot}^{-1}\sum_{i\in s_d}w_{di}e_{di}$.
Moreover, under the unit-level model \eqref{UnitlinearMixModel}, the model errors in \eqref{AreaLevelModel} satisfy $\bar{e}_{dw} \stackrel{ind}{\sim} N(0,\psi_d(\sigma_e^2))$, where
\begin{equation}
\psi_d(\sigma_e^2)=\sigma_{e}^2w_{d\cdot}^{-2}\sum_{i\in s_d}w_{di}^2, \quad d=1,\ldots,D.
\label{varedw}
\end{equation}
The aggregated model \eqref{AreaLevelModel} has an important advantage over the usual FH model, which is that all of the error variances $\psi_d(\sigma_e^2)$, $d=1,\ldots,D$, depend on a common parameter $\sigma_{e}^2$, for which a consistent estimator as $D\to\infty$ may be obtained from the available area- or unit-level data for the $D$ areas. In contrast, the usual FH model assumes that the error variances $\psi_d$, $d=1,\ldots,D$, are known, and they are often replaced with ``direct'' estimators $\psi_{d0}=\widehat{\var }_{\pi}(\hat\mu_d^{DIR}|\mu_d)$, $d=1,\ldots,D$, obtained from the $n_d$ unit-level survey observations from area $d$. Note that fitting the aggregated model \eqref{AreaLevelModel} requires the same data sources as those used to fit the usual FH model.

\cite{prasad1999robust} defined the Pseudo BLUP of $\mu_d=\bar{\boldsymbol{X}}_d'\boldsymbol{\beta}+u_d$ as the BLUP of $\mu_d$ under the aggregated model \eqref{AreaLevelModel}. The Pseudo BP of $\mu_d=\bar{\boldsymbol{X}}_d'\boldsymbol{\beta}+u_d$ under normality is
\begin{equation}
	\tilde{\mu}_{d}^{YR}(\bbeta,\sigma_u^2,\sigma_e^2) = \gamma_d(\sigma_u^2,\sigma_e^2)\left[\bar{y}_{dw}+(\bar{\boldsymbol{X}}_d - \bar{\boldsymbol{x}}_{dw})'\boldsymbol{\beta}\right] + \left(1-\gamma_d(\sigma_u^2,\sigma_e^2)\right)\bar{\boldsymbol{X}}_d'\boldsymbol{\beta},
	\label{pseudoBP}
\end{equation}
where now $\gamma_d(\sigma_u^2,\sigma_e^2)=\sigma_{u}^{2}/(\sigma_{u}^{2} +\psi_d(\sigma_e^2))$, for $\psi_d(\sigma_e^2)$ given in \eqref{varedw}.

\cite{prasad1999robust} proposed to estimate $\bbeta$ in the aggregated area-level model \eqref{AreaLevelModel} using the aggregated data $\{(\bar{y}_{dw},\bar{\boldsymbol{x}}_{dw}); \ d=1,\ldots,D\}$. However, $\sigma_u^2$ and $\sigma_e^2$ were estimated using a moments method based on the unit-level data. Then \cite{prasad1999robust} defined the Pseudo EBLUP of $\mu_d$ by replacing the unknown $\bbeta$, $\sigma_u^2$ and $\sigma_e^2$, with the resulting estimates in \eqref{pseudoBP}.
Later, \cite{you2002pseudo} noted that, since $n\geq D$, a more efficient Pseudo EBLUP is obtained by estimating $\bbeta$ in the original unit-level model \eqref{UnitlinearMixModel} using the unit-level data $\{(y_{di},\boldsymbol{x}_{di},w_{di}); \ i\in s_d,\, d=1,\ldots, D\}$ as follows,
\begin{equation}
	\tilde{\boldsymbol{\beta}}_U (\sigma_u^2,\sigma_e^2)= \left[\sum_{d=1}^{D}\sum_{i\in s_d}w_{di}\boldsymbol{x}_{di}\left(\boldsymbol{x}_{di}-\gamma_d(\sigma_u^2,\sigma_e^2)\bar{\boldsymbol{x}}_{dw}\right)'\right]^{-1} \sum_{d=1}^{D}\sum_{i\in s_d} w_{di}\left(\boldsymbol{x}_{di}-\gamma_d(\sigma_u^2,\sigma_e^2)\bar{\boldsymbol{x}}_{dw}\right)y_{di}.
	\label{BetapseudoEBLUP}
\end{equation}
Estimators $\hat{\sigma}_{u,U}^{2}$ and $\hat{\sigma}_{e,U}^{2}$ of $\sigma_u^2$ and $\sigma_e^2$, respectively, were also obtained using the method of Henderson III (H3) based on the unit-level data. \cite{you2002pseudo} then defined another Pseudo EBLUP of $\mu_d$ as
$\hat{\mu}_{d}^{YR}=\tilde{\mu}_{d}^{YR}(\hat\bbeta_U,\hat\sigma_{u,U}^2,\hat\sigma_{e,U}^2)$, where $\hat\bbeta_U=\tilde{\boldsymbol{\beta}}_U (\hat\sigma_{u,U}^2,\hat\sigma_{u,U}^2)$.

\cite{you2002pseudo} noted that, if survey weights $w_{di}$, $i\in s_d$, are calibrated so that $\sum_{i\in s_d}w_{di}=N_d$, the Pseudo EBLUP $\hat{\mu}_{d}^{YR}$ obtained considering intercept (i.e., with $x_{di1}=1$) automatically satisfies a benchmarking property. More specifically, when adding the Pseudo EBLUPs of the area totals over all the areas of the population, we obtain the survey regression estimator of the population total $Y=\sum_{d=1}^D\sum_{i=1}^{N_d}y_{di}$, that is,
\begin{equation}
	\sum_{d=1}^{D} N_d \hat{\mu}_{d}^{YR} = \hat{Y}_{w} + (\boldsymbol{X} - \hat{\boldsymbol{X}}_{w})'\hat{\boldsymbol{\beta}}_U,
\end{equation}
where $\hat{Y}_{w}=\sum_{d=1}^{D}\sum_{i\in s_d}w_{di}y_{di}$ and $\hat{\boldsymbol{X}}_{w}=\sum_{i\in s_d}w_{di}\boldsymbol{x}_{di}$ are the survey-weighted direct estimators of the population totals $Y$ and $\boldsymbol{X}=\sum_{d=1}^D\sum_{i=1}^{N_d}\boldsymbol{x}_{di}$, respectively.

\cite{you2002pseudo} approximated the MSE of the Pseudo EBLUP $\hat{\mu}_{d}^{YR}$ when $\sigma_u^2$ and $\sigma_e^2$ are estimated by H3 method, with
\begin{equation}
	\mbox{MSE}_{YR}(\hat{\mu}_{d}^{YR}) = g_{1dw}(\sigma_{u}^{2}, \sigma_{e}^{2}) + g_{2dw}(\sigma_{u}^{2}, \sigma_{e}^{2}) + g_{3dw}(\sigma_{u}^{2}, \sigma_{e}^{2}),
	\label{MSEPEBLUP}
\end{equation}	
where
\begin{equation*}
	g_{1dw}(\sigma_{u}^{2}, \sigma_{e}^{2})  = \left(1-\gamma_d(\sigma_u^2,\sigma_e^2)\right)\sigma_{u}^2,
\end{equation*}
\begin{equation*}
	g_{2dw}(\sigma_{u}^{2}, \sigma_{e}^{2})  = \left(\bar{\boldsymbol{X}}_d - \gamma_d(\sigma_u^2,\sigma_e^2)\bar{\boldsymbol{x}}_{dw} \right)'\Phi_w\left(\bar{\boldsymbol{X}}_d - \gamma_d(\sigma_u^2,\sigma_e^2)\bar{\boldsymbol{x}}_{dw}\right),
\end{equation*}
\begin{equation}
	g_{3dw}(\sigma_{u}^{2}, \sigma_{e}^{2})  = \gamma_d(\sigma_u^2,\sigma_e^2)\left(1-\gamma_d(\sigma_u^2,\sigma_e^2)\right)^2\sigma_e^{-4}\sigma_u^{-2}h(\sigma_{u}^{2}, \sigma_{e}^{2}),
 \label{g3dw}
\end{equation}
where $h(\sigma_{u}^{2}, \sigma_{e}^{2})$ and $\Phi_w$ and are respectively given in (7) and (16) of \cite{you2002pseudo}. For the same fitting method, they provided the following second-order unbiased MSE estimator,
\begin{equation}
\mbox{mse}_{YR}(\hat{\mu}_d^{YR}) =  g_{1dw}(\hat{\sigma}_{u,U}^2, \hat{\sigma}_{e,U}^{2}) + g_{2dw}(\hat{\sigma}_{u,U}^2, \hat{\sigma}_{e,U}^{2}) + 2g_{3dw}(\hat{\sigma}_{u,U}^2, \hat{\sigma}_{e,U}^{2}).
\label{msePEBLUP}
\end{equation}

\section{Unification of unit-level and area-level models}\label{sec4}

This section shows that aggregation of BHF unit-level model using the survey weights leads to a FH model, provided that the survey weights are calibrated to the known population totals of the auxiliary variables at the area-level.

Calibration estimators were defined by \cite{deville1992} as expansion estimators based on new weights $w_{di}^C$, $i\in s_d$, obtained by minimizing $\sum_{i\in s_d}d(w_{di},w_{di}^C)$, where $d(\cdot,\cdot)$ is a pseudo-distance, subject to certain calibration restrictions.
In this section, we consider that $w_{di}^C$, $i\in s_d$, are obtained by calibration of the original survey weights $w_{di}$, $i\in s_d$, to the known population totals of the auxiliary variables $\boldsymbol{X}_d=\sum_{i=1}^{N_d}\boldsymbol{x}_{di}$ in area $d$, that is, satisfying the calibration restrictions $\sum_{i\in s_d} w_{di}^C\boldsymbol{x}_{di}=\boldsymbol{X}_d$.
Taking the first element of $\boldsymbol{x}_{di}$ as one, the first calibration restriction is actually $\sum_{i\in s_d} w_{di}^C=N_d$.

The most common type of calibration is linear calibration, obtained using the chi-squared pseudo-distance $d(w_{di},w_{di}^C)=(w_{di}-w_{di}^C)^2/w_{di}$. Actually, linear calibration of the area survey weights $w_{di}$, $i\in s_d$, to the area totals $\boldsymbol{X}_d$ leads to the well-known GREG estimator of the population total for area $d$, $\hat Y_d^{GREG}=\sum_{i\in s_d} w_{di}^Cy_{di}$.

Let $\bar{\boldsymbol{x}}_{dw}^C=N_d^{-1}\sum_{i\in s_d} w_{di}^C\boldsymbol{x}_{di}$ and $\bar{y}_{dw}^C=N_d^{-1}\sum_{i\in s_d} w_{di}^Cy_{di}$ be the weighted area means of the response variable and the auxiliary variables, respectively, using the calibrated weights.
Note that the calibration restrictions are $\bar{\boldsymbol{x}}_{dw}^C=\bar{\boldsymbol{X}}_d$. Consequently, the model \eqref{AreaLevelModel}, obtained by aggregation of the unit-level BHF model \eqref{UnitlinearMixModel}, but using now the calibrated survey weights $w_{di}^C$, is a FH model as in \eqref{mixMod}, based on the direct estimator $\hat\mu_d^{DIR}=\bar{y}_{dw}^{C}$. More specifically, the aggregated model \eqref{AreaLevelModel} based on the calibrated weights $w_{di}^C$ reads
\begin{equation}
	\bar{y}_{dw}^C= \bar{\boldsymbol{X}}_{d}'\boldsymbol{\beta}+ u_d + \bar{e}_{dw}^{C}, \quad \bar{e}_{dw}^{C} \stackrel{ind}{\sim} N(0, \psi_{d}^C(\sigma_{e}^2)), \quad d=1,\ldots,D.
	\label{NewFH}
\end{equation}
with error variances now given by
\begin{equation}
\psi_d^C(\sigma_{e}^2)=\sigma_{e}^2N_d^{-2}\sum_{i\in s_d} (w_{di}^C)^2, \quad d=1,\ldots,D.
\label{psidC}
\end{equation}

We have seen that survey-weighted aggregation of the unit-level BHF model using the calibrated weights $w_{di}^C$, $i\in s_d$, leads to a FH model. However, as already said, the aggregated FH model \eqref{NewFH}, obtained assuming that the unit-level BHF model holds, has an advantage over the usual FH model in \eqref{mixMod}. In the usual FH model, error variances $\psi_{d}$, $d=1,\ldots,D$ are assumed to be known but, in practice, they need to be estimated and are often replaced with the unstable estimators $\psi_{d0}=\widehat{\var}_\pi(\hat{\mu}_{d}^{DIR}|\mu_{d})$ , $d=1,\ldots,D$. Moreover, the uncertainty due to the estimation of $\psi_d$ is not accounted for in the usual MSE estimators of $\hat\mu_d^{FHD}$.

In contrast, in the aggregated area-level FH model \eqref{NewFH} obtained with the calibrated weights $w_{di}^C$, error variances $\psi_d^C(\sigma_{e}^2)$, $d=1,\ldots,D$, given in \eqref{psidC}, are specified in terms of a single parameter $\sigma_e^2$. This parameter $\sigma_e^2$ may be estimated, similarly as in the Pseudo EBLUP, either with the unit-level survey data by fitting the original BHF model, or by fitting the aggregated FH model \eqref{NewFH} to the corresponding area aggregates, when the analyst has no access to the original unit-level survey data.

It turns out that, when using the calibrated weights $w_{di}^C$, $i\in s_d$, the Pseudo BP $\tilde\mu_d^{YR}(\bbeta,\sigma_u^2,\sigma_e^2)$ of $\mu_d$ given in \eqref{pseudoBP} based on the unit-level BHF model, equals the BP $\tilde{\mu}_d(\bbeta,\sigma_u^2,\bpsi)$ of $\mu_d$ given in \eqref{bpFayModel2}, based on the FH model \eqref{NewFH}.
This predictor, obtained by unified area-level and unit-level models and referred to as ``unified'' predictor, is given by
\begin{equation}
\tilde{\mu}_d^{U}(\bbeta,\sigma_u^2,\sigma_e^2) =
\gamma_d^{C}(\sigma_u^2,\sigma_{e}^2)\bar{y}_{dw}^{C} + \left(1-\gamma_d^{C}(\sigma_u^2,\sigma_{e}^2)\right)\bar{\boldsymbol{X}}_d'\boldsymbol{\beta},\quad \gamma_d^{C}(\sigma_u^2,\sigma_{e}^2) = \frac{\sigma_{u}^2}{\sigma_{u}^2 + \psi_d^C(\sigma_{e}^2)}.
\label{UnifiedModel}
\end{equation}

It holds that $\gamma_d^{C}(\sigma_u^2,\sigma_{e}^2)\to 1$ as $n_d\to\infty$, provided that $w_{di}^C>0$, for all $i\in s_d$, and $\max_{i\in s_d}(w_{di}^C/N_d)=O(n_d^{-1})$, as proven in the Appendix. Hence, in that case, under the same conditions under which $\bar{y}_{dw}^{C}$ is design-consistent as $n_d\to\infty$, for known $\bbeta$, $\sigma_u^2$ and $\sigma_e^2$, the unified predictor $\tilde{\mu}_d^{U}(\bbeta,\sigma_u^2,\sigma_e^2)$ given in \eqref{UnifiedModel} is also design consistent as $n_d\to\infty$.

Empirical versions of the unified predictor in \eqref{UnifiedModel} may be obtained, similarly as the Pseudo EBLUP, either by fitting the original BHF model using the available unit-level survey data, or by fitting the aggregated FH model \eqref{NewFH} using the aggregated area-level data. In the first case, we can use the estimator of $\bbeta$ given in \eqref{BetapseudoEBLUP}, but using the calibrated weights $w_{di}^C$ rather than the original weights $w_{di}$, denoted
$\tilde{\boldsymbol{\beta}}_{w}^C(\sigma_u^2,\sigma_e^2)$. Consistent estimators $\hat{\sigma}_{u,U}^2$ and $\hat{\sigma}_{e,U}^2$ of $\sigma_{u}^2$ and $\sigma_{e}^2$, respectively, as $D\to\infty$, may be obtained by ML, REML or H3 methods, applied to the original BHF model. We define an empirical unified (EU) predictor as $\hat{\mu}_d^{U}=\tilde{\mu}_d^{U}(\hat\bbeta_U^C,\hat\sigma_{u,U}^2,\hat\sigma_{e,U}^2)$, for $\hat\bbeta_U^C=\tilde{\boldsymbol{\beta}}_U^C(\hat\sigma_{u,U}^2,\hat\sigma_{e,U}^2)$.

As proven in the Appendix, the EU predictors of the area totals add up to the expansion estimator of the population total $Y$ based on the calibrated weights, $\hat{Y}_{w}^{C}=\sum_{d=1}^D\sum_{i\in s_d}w_{di}^C y_{di}$, that is,
\begin{equation}
\sum_{d=1}^{D} N_d \hat{\mu}_d^{U} = \hat{Y}_{w}^{C}.
\label{selfbenchmarking}
\end{equation}

Now consider that the analyst is not given access to the unit-level survey data, only to the aggregated area-level data $\{(\bar{y}_{dw}^{C},\bar{\boldsymbol{X}}_d,W_{2d});\ d=1,\ldots,D\}$, where $W_{2d}=\sum_{i\in s_d}(w_{di}^C)^2$. The regression coefficients $\bbeta$ under the aggregated FH model \eqref{NewFH} may be estimated with the usual WLS estimator, as follows
\begin{equation}
	\tilde{\boldsymbol{\beta}}_A^C(\sigma_{u}^2,\sigma_{e}^2) = \left(\sum_{d=1}^{D} \gamma_d^{C}(\sigma_{u}^2,\sigma_{e}^2)\bar{\boldsymbol{X}}_d\bar{\boldsymbol{X}}_d' \right)^{-1} \sum_{d=1}^{D}\gamma_d^{C}(\sigma_{u}^2,\sigma_{e}^2) \bar{\boldsymbol{X}}_d \bar{y}_{dw}^{C},
\end{equation}
where the subscript $A$ now means that only area-level data are used.
Consistent estimators $\hat{\sigma}_{u,A}^2$ and $\hat{\sigma}_{e,A}^2$ of $\sigma_{u}^2$ and $\sigma_{e}^2$ respectively, as $D\to\infty$, may be also obtained based on the same area-level data, either by ML, REML or a method of moments. Then, a second EU predictor of $\mu_d$ is obtained by replacing  the model parameters in the unified predictor with these estimates, that is, as
$\hat{\mu}_d^{UA}=\tilde{\mu}_d^{U}(\hat{\boldsymbol{\beta}}_A^C,\hat{\sigma}_{u,A}^2,\hat{\sigma}_{e,A}^2)$, for $\hat{\boldsymbol{\beta}}_A^C=\tilde{\boldsymbol{\beta}}_A^C(\hat{\sigma}_{u,A}^2,\hat{\sigma}_{e,A}^2)$.

Advantages of the Pseudo EBLUP $\hat\mu_d^{YR}$ and the two EU predictors $\hat\mu_d^{U}$ or $\hat\mu_d^{UA}$ over the usual EBLUPs based on the FH model is that the specification of error variances in terms of a single parameter $\sigma_e^2$ allows us to find consistent estimators $\hat\psi_d^C=\hat\sigma_{e}^2N_d^{-2}\sum_{i\in s_d} (w_{di}^C)^2$, as $D\to\infty$, for the model error variances $\psi_d^C(\sigma_e^2)$, $d=1,\ldots,D$.
This, in turn, allows us to propose consistent parametric bootstrap MSE estimators that incorporate the uncertainty due to the estimation of the error variances, for general fitting procedures providing consistent estimators of the model parameters under the unit-level BHF model \eqref{UnitlinearMixModel}.

\section{FH model with new specification of error variances}

The agreement of the unit-level BHF and the area-level FH models holds only when considering direct estimators $\hat\mu_d^{DIR}=\bar y_{dw}^C$ that use the calibrated survey weights $w_{di}^C$, $i\in s_d$. On the other hand, the FH model is not tied to this particular direct estimator. However, the only theoretical justification of the FH model assumptions that we find is actually through aggregation of the unit-level model.
Still, calibration at the area level is not compulsory and one may consider the direct estimator $\hat{\mu}_d^{DIR}=\bar{y}_{dw}$ based on the original weights $w_{di}$, $i\in s_d$, which might be previously calibrated at a larger aggregation level.

We can still assume that the unit-level BHF model \eqref{UnitlinearMixModel} holds. Then, according to the aggregated model \eqref{AreaLevelModel}, the variance of $\hat{\mu}_d^{DIR}=\bar{y}_{dw}$ given $\mu_d=\bar{\boldsymbol{X}}_d'\boldsymbol{\beta}+u_d$, is given by
\begin{equation}
\var(\hat{\mu}_d^{DIR}|\mu_d)=\sigma_{e}^2w_{d\cdot}^{-2}\sum_{i\in s_d}w_{di}^2\triangleq \psi_d(\sigma_{e}^2), \quad d=1,\ldots,D.
\label{varydw}
\end{equation}
In this section, we consider the original FH model \eqref{mixMod} using $\hat{\mu}_d^{DIR}=\bar{y}_{dw}$, but with error variances $\psi_d=\psi_d(\sigma_e^2)$ as specified in \eqref{varydw}, depending on a single parameter $\sigma_e^2$. This parameter can be estimated using the area-level data from the $D$ areas. This approach accounts for the sampling design through $\psi_d(\sigma_e^2)$, but avoids assuming that $\psi_d$ are known and the application of smoothing procedures for $\psi_{d0}$, $d=1,\ldots,D$.

Using the specification of error variances in \eqref{varydw} in the FH model, the BP of $\mu_d=\bar{\boldsymbol{X}}_d'\bbeta + u_d$ under that model with normality is the same as in \eqref{bpFayModel2}, but we now write it making explicit the dependence on a single variance parameter $\sigma_e^2$, as
\begin{equation}
\tilde{\mu}_d(\bbeta,\sigma_u^2,\sigma_e^2) =
\gamma_d(\sigma_u^2,\sigma_e^2)\bar{y}_{dw} + \left(1-\gamma_d(\sigma_u^2,\sigma_e^2)\right)\bar{\boldsymbol{X}}_d'\boldsymbol{\beta}.
\label{FHAmodel}
\end{equation}

Considering only the aggregated data, the BLUP of $\mu_d = \bar{\boldsymbol{X}}_d'\bbeta + u_d$ is obtained by replacing the true $\bbeta$ in the BP $\tilde{\mu}_d(\bbeta,\sigma_u^2,\sigma_e^2)$ by the WLS estimator as in \eqref{BetaFH}, given now by
\begin{equation}
\tilde{\boldsymbol{\beta}}_{FHA}(\sigma_u^2,\sigma_e^2) = \left( \sum_{d=1}^{D}\gamma_d(\sigma_u^2,\sigma_e^2) \bar{\boldsymbol{X}}_d\bar{\boldsymbol{X}}_d' \right)^{-1} \sum_{d=1}^{D} \gamma_d(\sigma_u^2,\sigma_e^2)\bar{\boldsymbol{X}}_d \bar{y}_{dw}.
\label{BetaFHA}
\end{equation}

Consistent estimators $\hat\sigma_{u,A}^2$ and $\hat\sigma_{e,A}^2$ of $\sigma_u^2$ and $\sigma_e^2$, respectively, as $D\to\infty$, may be obtained by ML, REML or H3 methods applied to the FH model \eqref{mixMod} with $\psi_d=\psi_d(\sigma_e^2)$ given in \eqref{varydw}. Based on the resulting estimators, we define an empirical best predictor as $\hat{\mu}_d^{FHA}=\tilde{\mu}_d(\hat{\bbeta}_{FHA},\hat{\sigma}_{u,A}^2,\hat{\sigma}_{e,A}^2)$, where $\hat{\bbeta}_{FHA} = \tilde{\boldsymbol{\beta}}_{FHA}(\hat{\sigma}_{u,A}^2,\hat{\sigma}_{e,A}^2)$.

An asymptotic approximation to the MSE of $\hat{\mu}_d^{FHA}$ and its corresponding analytical estimator may be obtained following similar arguments as those used by \cite{prasad1990estimation}. In Section \ref{sec4.2} we propose a parametric bootstrap MSE estimator.

\section{MSE of PEBLUP and EU predictor}\label{MSEunified}

When using H3 method to estimate the variance components, the MSE of PEBLUP can be estimated by the analytical second-order unbiased estimator given in \eqref{msePEBLUP}. Similarly, using the same estimation method, an estimator for the MSE of the EU predictor $\hat{\mu}_d^{U}$ is obtained by replacing $\gamma_d(\sigma_u^2,\sigma_e^2)$ with $\gamma_d^{C}(\sigma_u^2,\sigma_e^2)$ and $\bar{\boldsymbol{x}}_{dw}$ with $\bar{\boldsymbol{x}}_{dw}^C$ in \eqref{msePEBLUP}. This estimator does account for the uncertainty due to the estimation of all the unknown model parameters, but is specific for H3 estimators of $\sigma_u^2$ and $\sigma_e^2$.

Below, we describe a parametric bootstrap procedure that provides  alternative estimators for the MSE of $\hat{\mu}_d^{YR}$ or $\hat{\mu}_d^{U}$, applicable to more general fitting methods. The resulting bootstrap MSE estimator is consistent for the true MSE as $D\to\infty$, at the same rate and under the same conditions as the consistency of the estimators of the parameters in the unit-level BHF model \eqref{UnitlinearMixModel}. This result can be proven using the method of imitation, similar to the approach followed by \cite{gonzalez2008bootstrap} for H3. The bootstrap procedure is outlined below for $\hat{\mu}_d^{YR}$. Replacing in this procedure $\hat{\mu}_d^{YR}$ with $\hat{\mu}_d^{U}$, based on the expansion estimator $\hat\mu_d^{DIR}=\bar y_{dw}^C$ that uses the calibrated survey weights $w_{di}^C$ rather than the original weights $w_{di}$, it yields a parametric bootstrap MSE estimator for $\hat{\mu}_d^{U}$.
\vspace{0.3 cm}\\
{\bf Parametric bootstrap for MSE of PEBLUP and EU predictor}
\begin{enumerate}
\item Fit BHF model \eqref{UnitlinearMixModel} to the unit-level survey data $\{(y_{di}, \boldsymbol{x}_{di}); \ i\in s_d,\, d=1,\ldots, D\}$, obtaining consistent estimators $\hat{\boldsymbol{\beta}}_U$, $\hat{\sigma}_{u,U}^{2}$ and $\hat{\sigma}_{e,U}^{2}$ of $\boldsymbol{\beta}$, $\sigma_{u}^{2}$ and $\sigma_{e}^{2}$, respectively.
\item For $b=1,\ldots,B$ with $B$ large:
\begin{enumerate}
	\item[(i)] Generate area effects and bootstrap errors for sampled units as
  $$
  u_{d}^{*(b)}\stackrel{iid}{\sim} N(0,\hat{\sigma}_{u,U}^{2}),\quad e_{di}^{*(b)}\stackrel{iid}{\sim} N(0, \hat{\sigma}_{e,U}^{2}),\quad i \in s_d,\ d=1,\ldots,D.
  $$
	\item[(ii)] With the generated area effects $u_{d}^{*(b)}$ from (i) and the fitted model parameters obtained in step 1, obtain the true bootstrap area means as
  $$
  \mu_d^{*(b)}= \bar{\boldsymbol{X}}_d'\hat{\boldsymbol{\beta}}_U + u_{d}^{*(b)},\quad d=1,\ldots,D.
  $$
	\item[(iii)] Calculate the bootstrap observations for the sample units as
  \begin{equation}\label{bootunitlevel}
  y_{di}^{*(b)}=\boldsymbol{x}_{di}'\hat{\boldsymbol{\beta}} + u_{d}^{*(b)} + e_{di}^{*(b)},\quad i \in s_d,\ d=1,\ldots,D.
  \end{equation}
	\item[(iv)] Fit the bootstrap unit-level BHF model \eqref{bootunitlevel}, obtaining bootstrap model parameter estimates $\hat{\sigma}_{u,U}^{2*(b)}$, $\hat{\sigma}_{e,U}^{2*(b)}$ and $\hat{\boldsymbol{\beta}}_U^{*(b)}$, and from them obtain $\hat{\mu}_d^{YR*(b)}$.
\end{enumerate}
\item A parametric bootstrap estimator of $\mbox {MSE}(\hat{\mu}_d^{YR})$ is then given by
\begin{equation}
\mbox{mse}_{PB}(\hat{\mu}_d^{YR}) = \frac{1}{B}\sum_{b=1}^{B}\left(\hat{\mu}_d^{YR*(b)} - \mu_d^{*(b)}\right)^{2}.
\label{msePB}
\end{equation}
\end{enumerate}

For the customary EBLUP under the FH model, $\hat\mu_d^{FHD}$, obtained with the direct estimator $\hat\mu_d^{DIR}=\bar{y}_{dw}^{C}$ based on the calibrated weights $w_{di}^C$, we propose in Section \ref{sec4.2} a consistent parametric bootstrap MSE estimator which, unlike $\mbox{mse}_{PR}(\hat{\mu}_{d}^{FHD})$, incorporates the uncertainty due to the estimation of $\psi_{d}$ with $\psi_{d0}=\widehat{\var}_\pi(\bar{y}_{dw}^{C}|\mu_{d})$, $d=1,\ldots,D$.

\section{MSE of EBLUP under FH model}\label{sec4.2}

The use of the direct estimator $\hat{\mu}_d^{DIR}=\bar{y}_{dw}^C$ based on the calibrated survey weights $w_{di}^C$ in FH model provides a specification of the error variances in terms of a common parameter $\sigma_e^2$ for all the areas. This specification allows us to estimate these variances consistently as $D\to\infty$, similarly to the other model parameters. The consistency of the parameter estimators used to generate new data under a bootstrap procedure yields consistency of the parametric bootstrap MSE estimator.

We propose here two different parametric bootstrap estimators for the MSE of the EBLUP $\hat\mu_d^{FHD}$ of $\mu_d$ based on the usual FH model \eqref{mixMod}. The second MSE estimator uses the Prasad-Rao analytical MSE estimator given in \eqref{msePR}, correcting this formula for the missing uncertainty based on parametric bootstrap.
\vspace{0.3 cm}\\
{\bf Parametric bootstrap for MSE of EBLUP under FH with calibrated weights}
\begin{enumerate}
	\item Obtain consistent estimators, as $D\to\infty$, $\hat{\boldsymbol{\beta}}$, $\hat{\sigma}_{u}^{2}$ and $\hat{\sigma}_{e}^{2}$ of $\boldsymbol{\beta}$, $\sigma_{u}^{2}$ and $\sigma_{e}^{2}$ respectively. They can be obtained either by fitting the unit-level BHF model \eqref{UnitlinearMixModel} to the unit-level survey data $\{(y_{di}, \boldsymbol{x}_{di}); \ i\in s_d,\, d=1,\ldots, D\}$, or by fitting the aggregated model \eqref{NewFH} using the area-level data $\{(\bar{y}_{dw}^{C},\bar{\boldsymbol{X}}_d);\ d=1,\ldots,D\}$.
	\item For $b=1,\ldots,B$ with $B$ large:
	\begin{enumerate}
	\item[(i)] Taking as true error variances $\hat{\psi}_{d}=\hat{\sigma}_{e}^2N_d^{-2}\sum_{i\in s_d} (w_{di}^C)^2$, $d=1,\ldots,D$, generate, independently, bootstrap area effects and sampling errors as
  $$
  u_{d}^{*(b)}\stackrel{iid}{\sim} N(0, \hat{\sigma}_{u}^{2}),\quad e_{d}^{*(b)}\stackrel{ind}{\sim} N(0, \hat{\psi}_{d}),\quad d=1,\ldots,D.
  $$
 \item[(ii)] With $u_{d}^{*(b)}$ generated in (i), obtain the true bootstrap area means as
  $$
  \mu_d^{*(b)}= \bar{\boldsymbol{X}}_d'\hat{\boldsymbol{\beta}} + u_{d}^{*(b)},\quad d=1,\ldots,D.
  $$
\item[(iii)] Compute the bootstrap direct estimators
  $$
  \hat{\mu}_d^{DIR*(b)}=\mu_d^{*(b)} + e_{d}^{*(b)}, d=1,\ldots,D.
  $$
\item[(iv)] Taking $\hat\mu_d^{DIR}=\bar{y}_{dw}^{C}$ and $\psi_{d0}=\widehat{\mbox{\var}}_{\pi}(\bar{y}_{dw}^{C}|\mu_d)$, fit the FH model \eqref{mixMod} to the area-level data $\{(\bar{y}_{dw}^{C},\bar{\boldsymbol{X}}_d,\psi_{d0});\ d=1,\ldots,D\}$, and calculate $\hat{\mu}_d^{FHD*(b)}$.
\end{enumerate}
\item A parametric bootstrap estimator of $\mbox{MSE}(\hat{\mu}_d^{FHD})$ is then
\begin{equation}
\mbox{mse}_{PB1}(\hat{\mu}_d^{FHD}) = \frac{1}{B}\sum_{b=1}^{B}\left(\hat{\mu}_d^{FHD*(b)} - \mu_d^{*(b)}\right)^{2}.
\label{mseB1}
\end{equation}
\end{enumerate}

We now propose an alternative MSE estimator based on correcting through a bootstrap procedure the Prasad-Rao analytical MSE estimator $\mbox{mse}_{PR}(\hat{\mu}_d^{FHD})$ given in \eqref{msePR}, by incorporating the uncertainty due to replacing $\psi_{d}$ with $\psi_{d0}=\widehat{\var}_\pi(\bar{y}_{dw}^{C}|\mu_{d})$.
The bootstrap procedure is the same as the previous one, but additionally, in step 2(iv), calculates the EBLUP based on the ``true'' bootstrap error variances, $\hat\bpsi=(\hat{\psi}_1,\ldots,\hat\psi_D)'$, $d=1,\ldots,D$, denoted as $\hat{\mu}_d^{FHT*(b)}=\hat{\mu}_d^{FH*(b)}(\hat\bpsi)$. Then, in step 3, it additionally calculates the parametric bootstrap estimator of the MSE of the EBLUP under FH model  based on the true error variances, $\hat{\mu}_d^{FHT}=\hat\mu_d^{FH}(\bpsi)$, as follows
\begin{equation}
\mbox{mse}_{PB}(\hat{\mu}_d^{FHT}) = \frac{1}{B}\sum_{b=1}^{B}\left(\hat{\mu}_d^{FHT*(b)} - \mu_d^{*(b)}\right)^{2}.
\label{mseBT}
\end{equation}
The MSE estimator \eqref{mseBT} does not account for the uncertainty due to the estimation of the error variances. Therefore, the difference between $\mbox{mse}_{PB1}(\hat{\mu}_d^{FHD})$ and $\mbox{mse}_{PB}(\hat{\mu}_d^{FHT})$ represents the uncertainty that we need to add to the PR estimator $\mbox{mse}_{PR}(\hat{\mu}_d^{FHD})$.
Defining $\Delta_{d}=\max\left\{0, \mbox{mse}_{PB1}(\hat{\mu}_d^{FHD}) - \mbox{mse}_{PB}(\hat{\mu}_d^{FHT}) \right\}$, another estimator of $\mbox{MSE}(\hat{\mu}_d^{FHD})$ is then
\begin{equation}
 \mbox{mse}_{PB2}(\hat{\mu}_d^{FHD}) = \mbox{mse}_{PR}(\hat{\mu}_d^{FHD}) + \Delta_{d}.
 \label{mseB2}
 \end{equation}

 When there is no access to the unit-level survey data, we may consider also the EU predictor with model parameters estimated based on the area-level data, $\hat{\mu}_d^{UA}=\tilde{\mu}_d^{U}(\hat{\boldsymbol{\beta}}_A^C,\hat\sigma_{u,A}^2,\hat\sigma_{e,A}^2)$. A consistent MSE estimator $\mbox{mse}_{PB1}(\hat{\mu}_d^{UA})$ may be obtained with the same parametric bootstrap procedure, but employing the area-level data in step 1 to obtain the model parameter estimators $\hat{\boldsymbol{\beta}}_A^C$, $\hat\sigma_{u,A}^2$ and $\hat\sigma_{e,A}^2$. Then, in step 2(iv), we calculate $\hat{\mu}_d^{UA*(b)}$ with estimated error variances $\hat\psi_{d,A}^C=\hat{\sigma}_{e,A}^2N_d^{-2}\sum_{i\in s_d} (w_{di}^C)^2$, $d=1,\ldots,D$, instead of $\hat{\mu}_d^{FHD*(b)}$.

Similarly, if the direct estimator $\hat\mu_d^{DIR}=\bar y_{dw}$ based on the original weights is used, an estimator of the MSE of $\hat\mu_d^{FHA}$, denoted as $\mbox{mse}(\hat\mu_d^{FHA})$, could be obtained using the same bootstrap procedure as above, by replacing $\hat\mu_d^{FHD}$ with $\hat\mu_d^{FHA}$ in step 2 (iv).

\section{Simulation experiments}\label{sec5}

This section describes Monte Carlo (MC) simulation experiments designed to compare the properties (bias, MSE) of the alternative estimators of the area means $\mu_d$ considered here and to study the performance of the proposed MSE estimators.

The first experiment compares the estimators of $\mu_d$ obtained using the calibrated weights: 1) Direct estimator $\hat{\mu}_d^{DIR}=\bar{y}_{dw}^{C}$; 2) EBLUP based on the FH model, $\hat{\mu}_{d}^{FHD}$, obtained setting $\hat{\mu}_d^{DIR}=\bar{y}_{dw}^{C}$ and $\psi_{d}=\widehat{\var}_\pi(\bar{y}_{dw}^{C}|\mu_{d})$; 3) Empirical unified predictor, $\hat{\mu}_{d}^{UA}$, where error variances are given by $\psi_{d}^C(\sigma_e^2)=\sigma_{e}^2N_d^{-2}\sum_{i\in s_d} (w_{di}^C)^2$ and $\sigma_e^2$ is estimated, along with $\sigma_u^2$ and $\bbeta$, with the area-level data; 4) Empirical unified predictor, $\hat{\mu}_d^{U}$, where error variances are also given by $\psi_{d}^C(\sigma_e^2)=\sigma_{e}^2N_d^{-2}\sum_{i\in s_d} (w_{di}^C)^2$, but estimating $\sigma_e^2$, $\sigma_u^2$ and $\bbeta$ with the survey data at the unit level.

In this simulation experiment, we consider a population of $N=250\,000$ units, distributed into $D=25$ areas, with $N_d=10\,000$ units in each area $d=1\ldots,D$. We consider two continuous auxiliary variables, $x_q$, $q=1,2$, whose values are generated as $x_{q,di}\sim \mbox{Gamma}(k_{qd},1)$ with $k_{1d}=5+3d/D$, $k_{2d}=2$, $i=1\ldots,N_d$, $d=1\ldots,D$. The vector of true regression coefficients is taken as $\boldsymbol{\beta}=(4, 0\mbox{.}5, -0\mbox{.}4)'$ and the variances of the area effects and errors are taken, respectively, as $\sigma_{u}^2=0\mbox{.}1^2$ and $\sigma_{e}^2=0\mbox{.}3^2$.

The sample $s$ is drawn by simple random sampling without replacement (SRSWOR), independently from each area $d$, with area sample sizes $n_d$ are taken from $\{3,5,10,15,50\}$, where each different value is repeated for 5 consecutive areas.
Then, the original survey weights $w_{di}=N_d/n_d$, $i\in s_d$, are calibrated for each area $d$ to $\boldsymbol{X}_d=(N_d,\bar X_{1d},\bar X_{2d})$, $d=1,\ldots,D$, where $\bar X_{qd}=\sum_{i=1}^{N_d}x_{q,di}$, $q=1,2$, using linear calibration.

The simulation experiment proceeds as follows. For each MC simulation $\ell=1,\ldots,L$, with $L=1\,000$, we generate a population of $y_{di}^{(\ell)}$ values, $i=1,\ldots,N_d$, $d=1,\ldots,D$, from the unit-level BHF model in \eqref{UnitlinearMixModel} with the above sizes, auxiliary variables and model parameters. From the generated population values, we calculate the true area means $\mu_d^{(\ell)}=\sum_{i=1}^{N_d}y_{di}^{(\ell)}$, $d=1,\ldots,D$. Then, taking the values $y_{di}^{(\ell)}$ for the sample units, $i\in s_d$, $d=1,\ldots,D$, we compute the four estimators of interest, namely $\hat{\mu}_{d}^{DIR(\ell)}$, $\hat{\mu}_{d}^{FHD(\ell)}$, $\hat{\mu}_d^{UA(\ell)}$ and $\hat{\mu}_{d}^{U(\ell)}$. The sample units, $s$, the auxiliary information $\boldsymbol{x}_{di}=(1,x_{1,di},x_{2,di})'$, $i=1,\ldots, N_d$, $d=1,\ldots, D$, as well as the calibrated weights, are held fixed across MC simulations.

For a generic estimator $\hat{\mu}_{d}$, we evaluate its performance in terms of relative bias (RB) and relative root MSE (RRMSE), approximated empirically as
\begin{equation*}
	\mbox{RB}(\hat{\mu}_{d}) =  \frac{ L^{-1} \sum_{\ell=1}^{A} (\hat{\mu}_{d}^{(\ell)} - \mu_d^{(\ell)})}{L^{-1} \sum_{\ell=1}^{L} \mu_{d}^{(\ell)}}, \quad \mbox{RRMSE}(\hat{\mu}_d) = \frac{ \sqrt{ L^{-1} \sum_{\ell=1}^{L} (\hat{\mu}_{d}^{(\ell)} - \mu_d^{(\ell)})^2} }{L^{-1} \sum_{\ell=1}^{L} \mu_{d}^{(\ell)}}.
\label{RMBd}
\end{equation*}
Average across domains of absolute RB (ARB) and RRMSE are also calculated as
\begin{equation*}
	\overline{\mbox{ARB}} = D^{-1} \sum_{d=1}^{D} |\mbox{RB}(\hat{\mu}_{d})|, \quad \overline{\mbox{RRMSE}} = D^{-1} \sum_{d=1}^{D} \mbox{RRMSE}(\hat{\mu}_{d}).
	\label{RB}
\end{equation*}

\begin{figure}[ht]%
	\centering	\includegraphics[width=160mm]{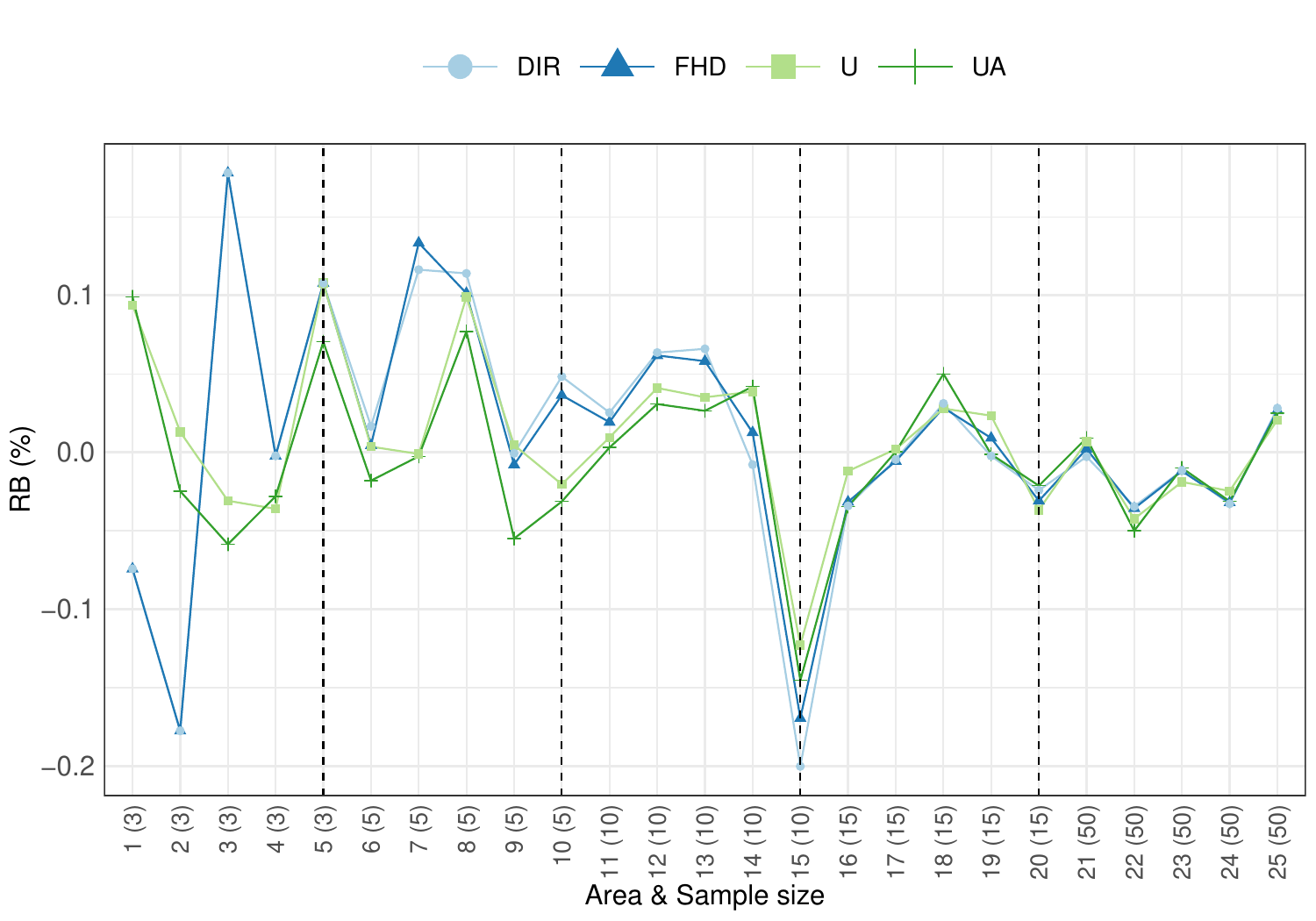}
	\caption{\% RB of FHD, UA and U estimators for each area. Sample sizes in parenthesis.}\label{fig:RBplot}
\end{figure}

Figure \ref{fig:RBplot} shows the percent RB of the four estimators, $\hat{\mu}_{d}^{DIR}$ (labelled DIR), $\hat{\mu}_{d}^{FHD}$ (labelled FHD), $\hat{\mu}_{d}^{UA}$ (labelled UA) and $\hat{\mu}_d^{U}$ (labelled U), for each domain $d=1,\ldots,D$ in the $x$-axis, with area sample sizes $n_d$ within parenthesis. This plot shows that the two EU predictors have similar RB. The same occurs for DIR and FHD, which have a larger absolute RB than the EU predictors, for small area sample sizes ($n_d<5$).

\begin{figure}[ht]%
	\centering	\includegraphics[width=160mm]{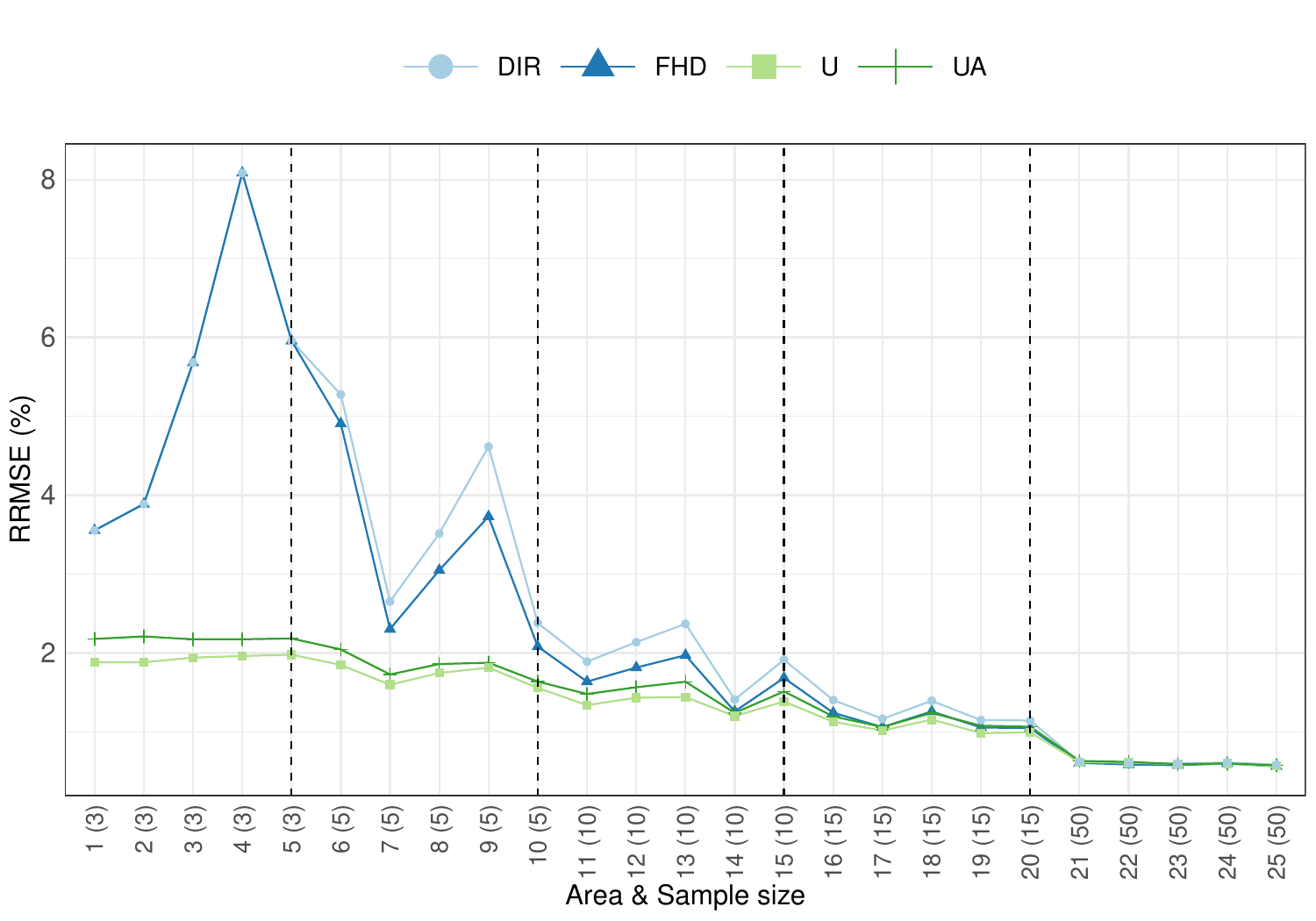}
	\caption{\% RRMSE of FHD, UA and U estimators for each area. Sample sizes in parenthesis.}\label{fig:RRMSEplot}
\end{figure}

Figure \ref{fig:RRMSEplot} displays the corresponding percent RRMSEs. This figure shows again a similar performance for the two EU predictors, although the one fitted using the more detailed unit-level data has slightly smaller RRMSEs. We can see that the RRMSE of FHD grows considerably for $n_d\leq 10$ and does not show much improvement compared with the direct estimator. Actually, it becomes as inefficient as the direct estimator for $n_d\leq 3$. This occurs because the variance estimators used in FHD are ``direct'', based on really small area sample sizes $n_d$. When using these inefficient variance estimators, the EBLUPs based on FH model become also inefficient. In contrast, the two EU predictors use consistent estimators of the error variances as $D\to\infty$. As expected, increasing the area sample size yields similar performance of all the estimators.

Table \ref{tab:Tab1} reports averages across the areas with the same sample size of ARB and RRMSE, for the three model-based estimators. In this table, FHD has an average RRMSE that is about twice that of UA for $n_d=3$, but this difference reduces as $n_d$ increases. U has the smallest average RRMSE for $n_d<50$, but for $n_d=50$, the average RRMSE and ARB values are very similar for the three estimators.

\begin{table*}[!t]
	\caption{Averages across areas with the same sample size of \% ARB and RRMSE, for FHD, UA and U estimators of $\mu_d$.}
	\tabcolsep=0pt
	\label{tab:Tab1}
	\begin{tabular*}{\textwidth}{@{\extracolsep{\fill}}lccccccccccc@{\extracolsep{\fill}}}
		\toprule%
		& \multicolumn{2}{@{}c@{}@{}}{$n_d=3$} & \multicolumn{2}{@{}@{}c@{}}{$n_d=5$} & \multicolumn{2}{@{}@{}c@{}}{$n_d=10$} & \multicolumn{2}{@{}@{}c@{}}{$n_d=15$} & \multicolumn{2}{@{}@{}c@{}}{$n_d=50$}\\
		\cline{2-3}\cline{4-5}\cline{6-7}\cline{8-9}\cline{10-11}
		Estimator & $\overline{\mbox{ARB}}$ & $\overline{\mbox{RRMSE}}$ & $\overline{\mbox{ARB}}$ & $\overline{\mbox{RRMSE}}$ & $\overline{\mbox{ARB}}$ & $\overline{\mbox{RRMSE}}$ & $\overline{\mbox{ARB}}$ & $\overline{\mbox{RRMSE}}$ & $\overline{\mbox{ARB}}$ & $\overline{\mbox{RRMSE}}$\\
		\midrule
		$\hat{\mu}_{d}^{FHD}$   & 0.11 & 5.43 & 0.06 & 3.21 & 0.06 & 1.67 & 0.02 & 1.13 & 0.02 & 0.59\\
		$\hat{\mu}_{d}^{UA}$  & 0.05 & 2.53 & 0.04 & 2.07 & 0.06 & 1.58 & 0.02 & 1.13 & 0.02 & 0.59\\
		$\hat{\mu}_d^{U}$      & 0.06 & 1.93 & 0.03 & 1.71 & 0.05 & 1.36 & 0.02 & 1.06 & 0.02 & 0.60\\
		\bottomrule
	\end{tabular*}
\end{table*}

The next MC simulation experiment evaluates the performance of the two available MSE estimators of the EU predictor $\hat{\mu}_d^{U}$, for finite number of areas ($D=25$): 1) The analytical second-order unbiased estimator $\mbox{mse}_{YR}(\hat{\mu}_d^{U})$, given by \eqref{msePEBLUP} with $\gamma_d(\sigma_u^2,\sigma_e^2)$ replaced by $\gamma_d^C(\sigma_u^2,\sigma_e^2)$ and $\bar{\boldsymbol{x}}_{dw}$ by $\bar{\boldsymbol{x}}_{dw}^C$; 2) The parametric bootstrap MSE estimator $\mbox{mse}_{PB}(\hat{\mu}_d^{U})$ proposed in Section \ref{MSEunified}. In this experiment, we consider area sample sizes $n_d$ in the sequence $\{5,10,15,20,25\}$, with each value repeated for five consecutive areas.

The true MSEs of the EU predictors were previously approximated as described above, but now with $L=100\,000$ MC replicates. Then, a new simulation is performed with $L=1\,000$ MC replicates, calculating in each replicate the analytical MSE estimator $\mbox{mse}_{YR}(\hat{\mu}_d^{U})$ and the parametric bootstrap estimator $\mbox{mse}_{PB}(\hat{\mu}_d^{U})$ given in \eqref{msePB} with $B=500$ bootstrap replicates.

\begin{figure}[ht]%
	\centering
	\includegraphics[width=160mm]{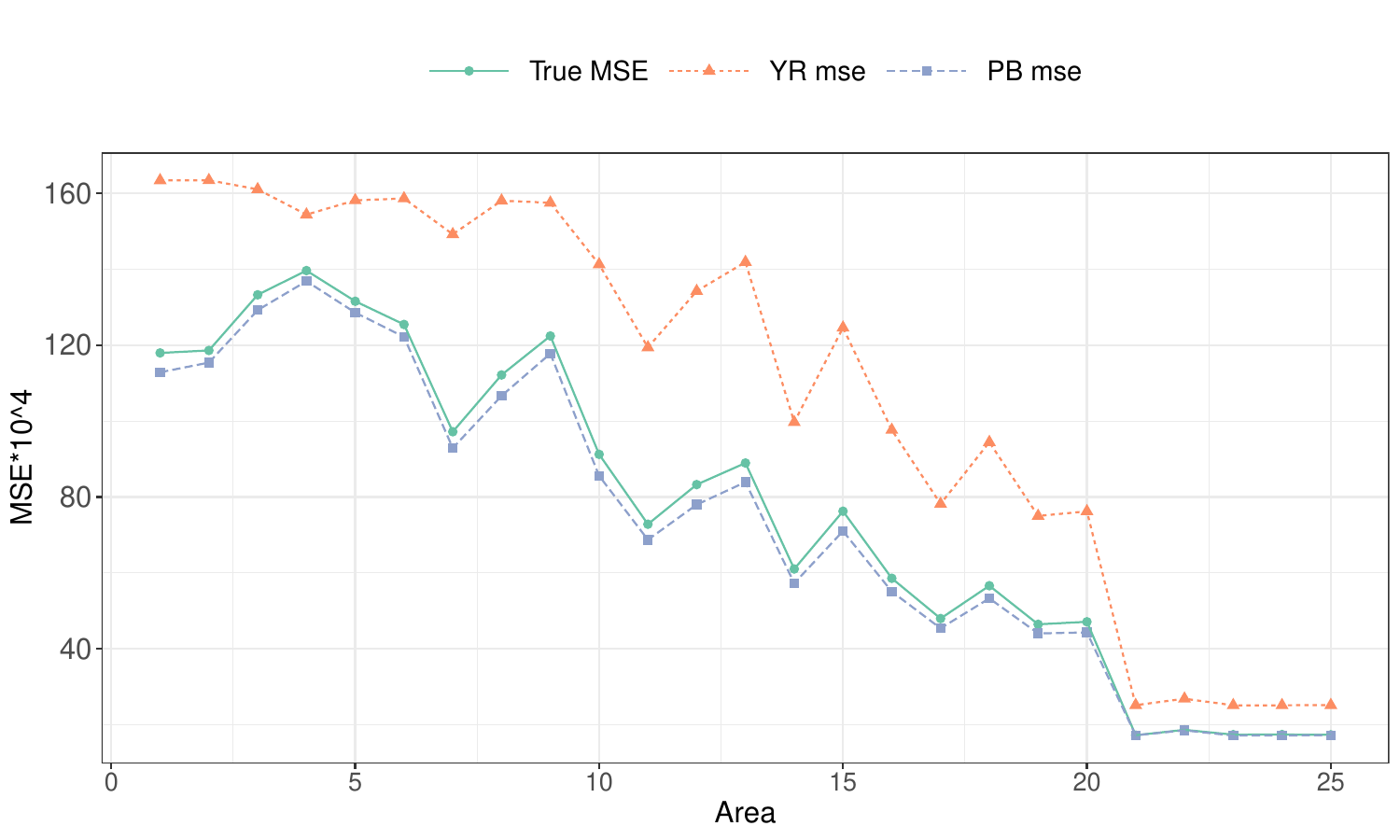}
	\caption{True MSEs of $\hat{\mu}^{U}_d$ and MC averages of YR and PB estimators for each area.}\label{fig:mseplot}
\end{figure}

Figure \ref{fig:mseplot} shows the true MSEs and the MC averages of the two MSE estimators, labelled YR and PB respectively, for each area. This plot shows that YR analytical MSE estimator overestimates the true MSEs for all the areas, but especially for those with $n_d\leq 20$. On the other hand, PB tracks very well the true MSEs for all the areas, performing clearly better than the analytical YR estimator, even for the areas with the largest sample sizes.

\begin{figure}[ht]%
	\centering
	\includegraphics[width=160mm]{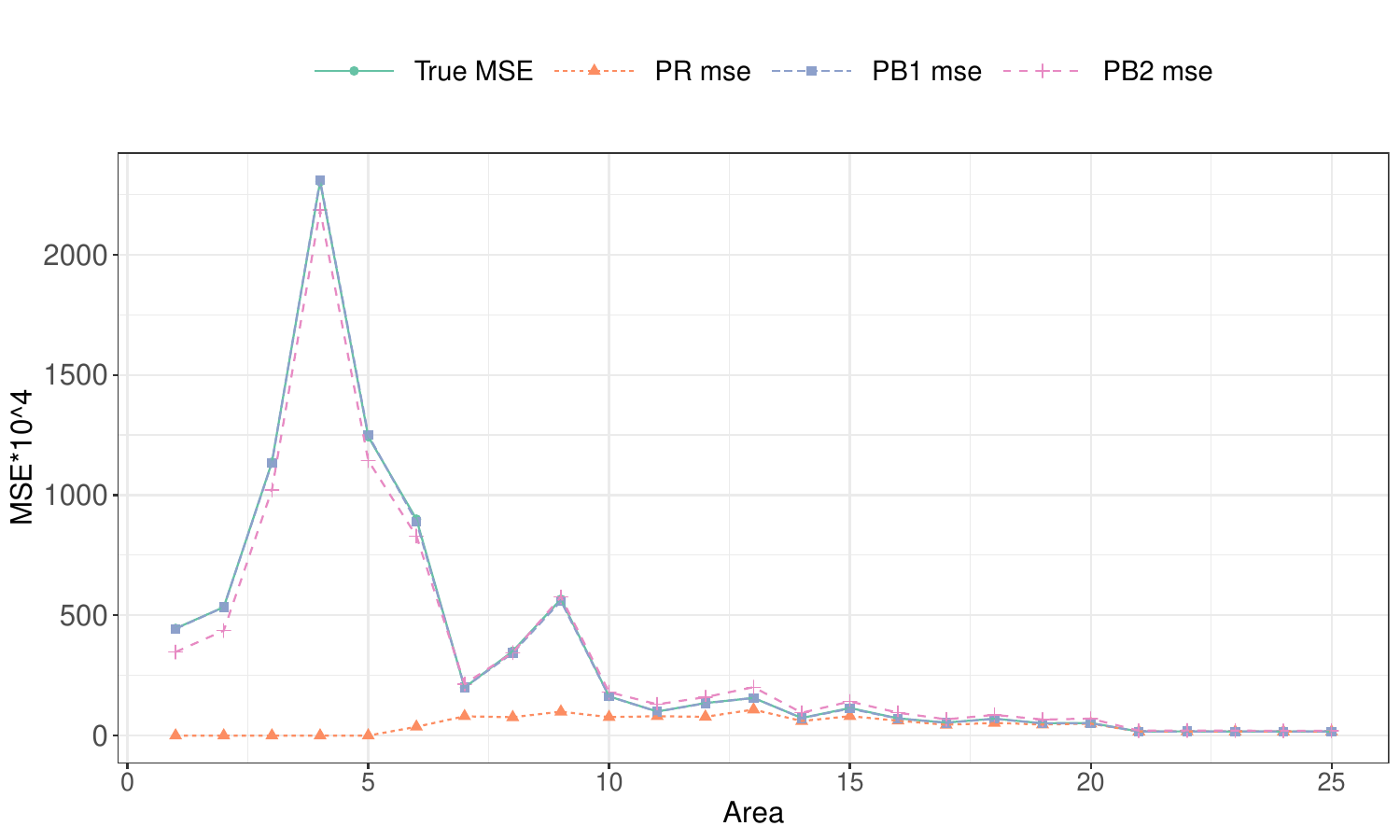}
	\caption{True MSEs of $\hat{\mu}^{FHD}_d$ and MC averages of PR, PB1 and PB2 estimators, for each area.}\label{fig:mseFHplot}
\end{figure}

Next we study the alternative estimators of the MSE of FHD. The true MSE of FHD was again previously approximated with $L=100\,000$ replicates. Figure \ref{fig:mseFHplot} shows the true MSEs, the means across MC simulations of the Prasad-Rao analytical MSE estimator $\mbox{mse}_{PR}(\hat{\mu}_d^{FHD})$ given in \eqref{msePR} (labelled PR mse) and of the two alternative parametric bootstrap MSE estimators $\mbox{mse}_{PB1}(\hat{\mu}_d^{FHD})$ and $\mbox{mse}_{PB2}(\hat{\mu}_d^{FHD})$ given in \eqref{mseB1} and \eqref{mseB2} (labelled PB1 mse and PB2 mse respectively).
This figure shows that PR analytical MSE estimator, which was not designed to account for the uncertainty due to the estimation of the error variances, underestimates seriously the true MSE of FHD estimator for the areas with $n_d\leq 10$, while the two proposed bootstrap estimators perform practically without error. For the areas with sample sizes greater than 10, the three MSE estimators perform similarly, becoming nearly equal for large $n_d$. Actually, PR mse is the analytical version of the parametric bootstrap MSE estimator of FHT, $\mbox{mse}_{PB}(\hat{\mu}_d^{FHT})$ given in \eqref{mseBT}, which assumes that the error variances $\psi_d$, $d=1,\ldots, D$ are known. None of them incorporate the error that arises from the estimation of these variances.

\begin{figure}[ht]%
	\centering
	\includegraphics[width=160mm]{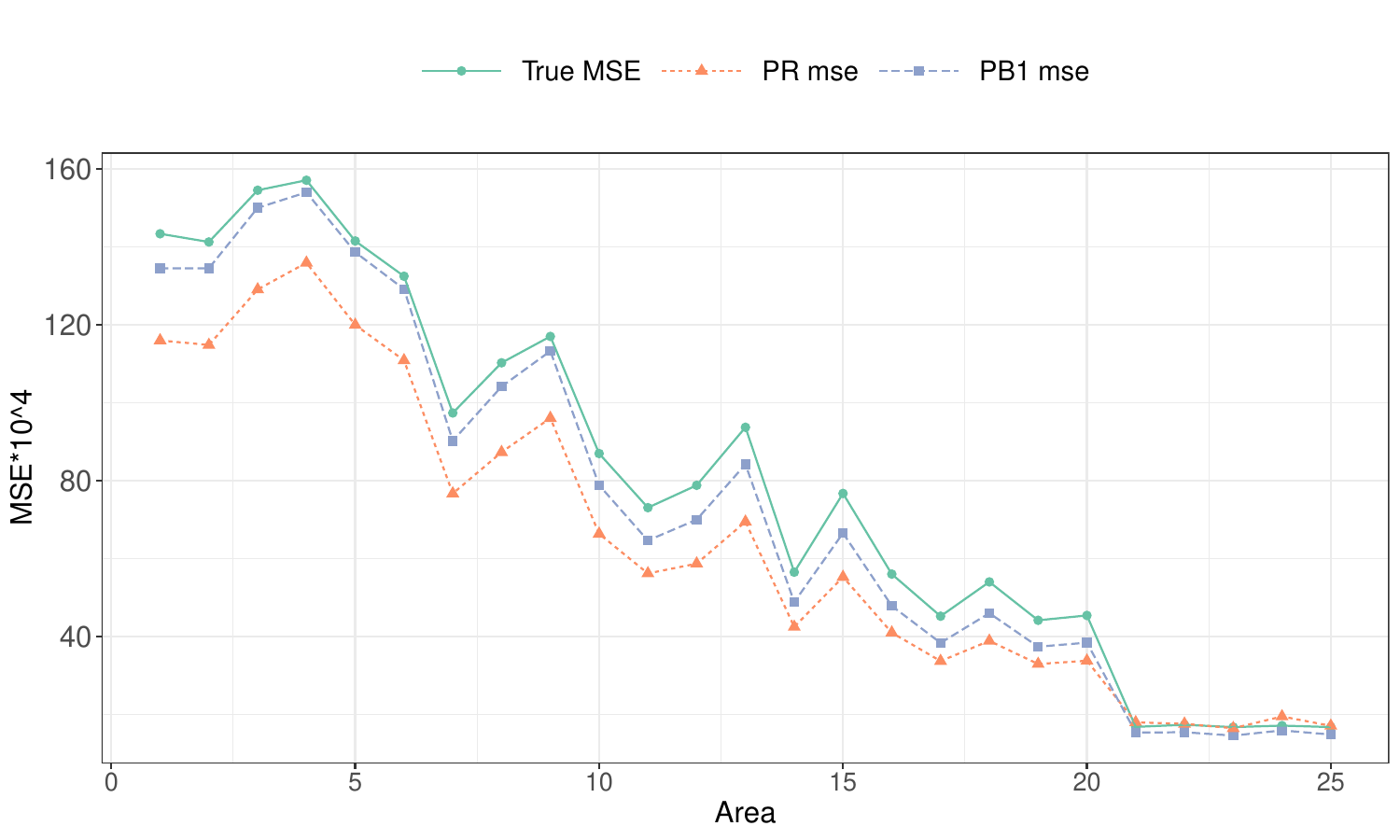}
	\caption{True MSEs of $\hat{\mu}_d^{UA}$ and MC averages of PR and PB1 estimators for each area.}\label{fig:mseFHTplot}
\end{figure}

We now investigate the performance of the MSE estimators for the EU predictor $\hat{\mu}_d^{UA}$ based on fitting the model with the area-level data. Figure \ref{fig:mseFHTplot} shows the true MSEs, and the MC averages of the Prasad-Rao analytical MSE estimator $\mbox{mse}_{PR}(\hat{\mu}_d^{UA})$ given in \eqref{msePR} (labelled PR mse) and of the parametric bootstrap MSE estimator $\mbox{mse}_{PB}(\hat{\mu}_d^{UA})$ obtained with the bootstrap procedure described in Section \ref{sec4.2} (labelled PB1 mse). This plot shows a better performance of PB1 estimator of the MSE, with PR mse clearly underestimating the true MSEs for $n_d\leq 20$. For $n_d\geq 20$, both MSE estimators are close to the true values.

We have obtained clear efficiency gains for the EU predictors of $\mu_d$, obtained when survey weights are calibrated to the known area counts and area totals of the auxiliary variables. Of course, this type of calibration is not compulsory when applying the FH model. However, the FH model has the problem of estimation of $\psi_d$, $d=1,\ldots, D$. Even though the equivalency between the EBLUP based on FH model and the Pseudo EBLUP based on BHF model holds only when using the calibrated survey weights, we also studied whether specifying the error variances $\psi_d$ in FH model to those variances obtained by aggregation of the unit-level BHF model still leads to efficiency gains compared to the usual FHD estimator obtained with the original survey weights.
Hence, in an additional simulation experiment, we compared the performance of the alternative estimators of $\mu_d$ obtained with the original weights $w_{di}$, $i\in s_d$: 1) Direct estimator $\hat\mu_d^{DIR}=\bar{y}_{dw}$; 2) EBLUP based on the FH model $\hat\mu_d^{FHD}$ that uses the direct estimator $\hat\mu_d^{DIR}=\bar{y}_{dw}$; 3) EBLUP based on the FH model $\hat\mu_d^{FHA}$ with error variances $\psi_d=\psi_d(\sigma_e^2)=\sigma_{e}^2w_{d\cdot}^{-2}\sum_{i\in s_d} w_{di}^2$, where $\sigma_{e}^2$ is estimated together with $\sigma_u^2$ and $\bbeta$ with area-level data; 4) Pseudo EBLUP (PEBLUP) $\hat\mu_d^{YR}$.

This simulation experiment was conducted under the same setup as before, with samples in each area drawn by SRSWOR, leading to constant survey weights $w_{di}=N_d/n_d$ for all $i\in s_d$, $d=1,\ldots,D$. Here the survey-weighted direct estimator $\hat\mu_d^{DIR}=\bar{y}_{dw}$ reduces to the area sample mean, $\bar{y}_{d}=n_d^{-1}\sum_{i\in s_d}y_{di}$, and the error variances resulting from aggregation in the unit-level model are given by $\psi_d(\sigma_{e}^2)=\sigma_{e}^2/n_d$, $d=1,\ldots,D$.

Figure \ref{fig:mseFHTplot_SRSWOR} in the Appendix shows RB and RRMSE of the four estimators, labelled as DIR, FHD, FHA and PEBLUP, respectively. This figure shows that, even using the original survey weights $w_{di}$ and fitting the FH model with area-level data only, FHA estimator obtained with the new specification of the error variances performs substantially better than FHD estimator. Moreover, the loss of efficiency with respect to PEBLUP, based on fitting the model with unit-level data, is small. Interestingly, FHD is even less efficient than the direct estimator in several areas. 

\section{Application}

In this section, we consider data from the Colombian \textit{Saber 11} test, an annual test conducted by the \textit{Colombian Institute for the evaluation of education} to measure the quality of formal high school education. We select the test for Calendar B students in 2018. This test evaluates Critical Reading, Mathematics, Social and Human Sciences, Natural Sciences and English. The data represent a census of the population of students and, from them, official education statistics are published. The data set includes also information about high schools, such as whether they are public or private or their geographical location, as well as socio-economic conditions of students.

Colombia is divided into $D=33$ administrative and political divisions, 32 departments and one capital district, hereafter called departments for simplicity. The department population sizes $N_d$ (number of students taking the test in that department) range from 268 to 87\,556. For this application, we select a SRSWOR of 1.5\% from each department. The total sample size is $n=8\,250$, where $n_d$ varies between 4 and 1\,313.

We are interested in estimating the mean $\mu_d$ of the Maths scores $y_{di}$, $i=1,\ldots,N_d$, for each of the $D=33$ departments, using as auxiliary information the number of rooms of the student's household, $x_{1,di}$.
We will compare the two alternative EU predictors $\hat{\mu}_{d}^{U}$ and $\hat{\mu}_{d}^{UA}$, with FHD estimator, $\hat{\mu}_{d}^{FHD}$. The MSE of $\hat{\mu}_{d}^{U}$ will be estimated with the parametric bootstrap estimator, $\mbox{mse}_{PB}(\hat{\mu}_{d}^{U})$, described in Section \ref{MSEunified}. For $\hat{\mu}_{d}^{FHD}$, we consider the PR analytical MSE estimator, $\mbox{mse}_{PR}(\hat{\mu}_{d}^{FHD})$, given in Section \ref{sec4.2}, and for $\hat{\mu}_{d}^{UA}$ we obtain the parametric bootstrap estimator obtained following the approach of Section \ref{sec4.2}, $\mbox{mse}_{PB1}(\hat{\mu}_{d}^{UA})$.

First of all, for each area $d=1,\ldots,33$, we calibrate the survey weights $w_{di}$, $i\in s_d$, to the known department count $N_d$ and the total $X_{1,d}=\sum_{i=1}^{N_d} x_{1,di}$, using linear calibration.
Now, since both EU predictors assume that BHF model holds for the Maths scores $y_{di}$, we check the model assumptions. Figure \ref{fig:residuals_model} in the Appendix shows a histogram and a normal Q-Q plot of unit-level residuals obtained from a REML fit of the model, given by $\hat{e}_{di}=y_{di}-\boldsymbol{x}_{di}'\hat{\boldsymbol{\beta}} - \hat{u}_d$, for $\hat{u}_d=\hat\gamma_d(\bar y_d-\bar \bx_d'\hat\bbeta)$ with $\hat\gamma_d=\hat\sigma_u^2/(\hat\sigma_u^2+\hat\sigma_e^2/n_d)$. This figure indicates that the normal distribution is tenable, showing only a mild deviation in the tails.

\begin{figure}[!h]%
	\centering
	\includegraphics[width=155mm]{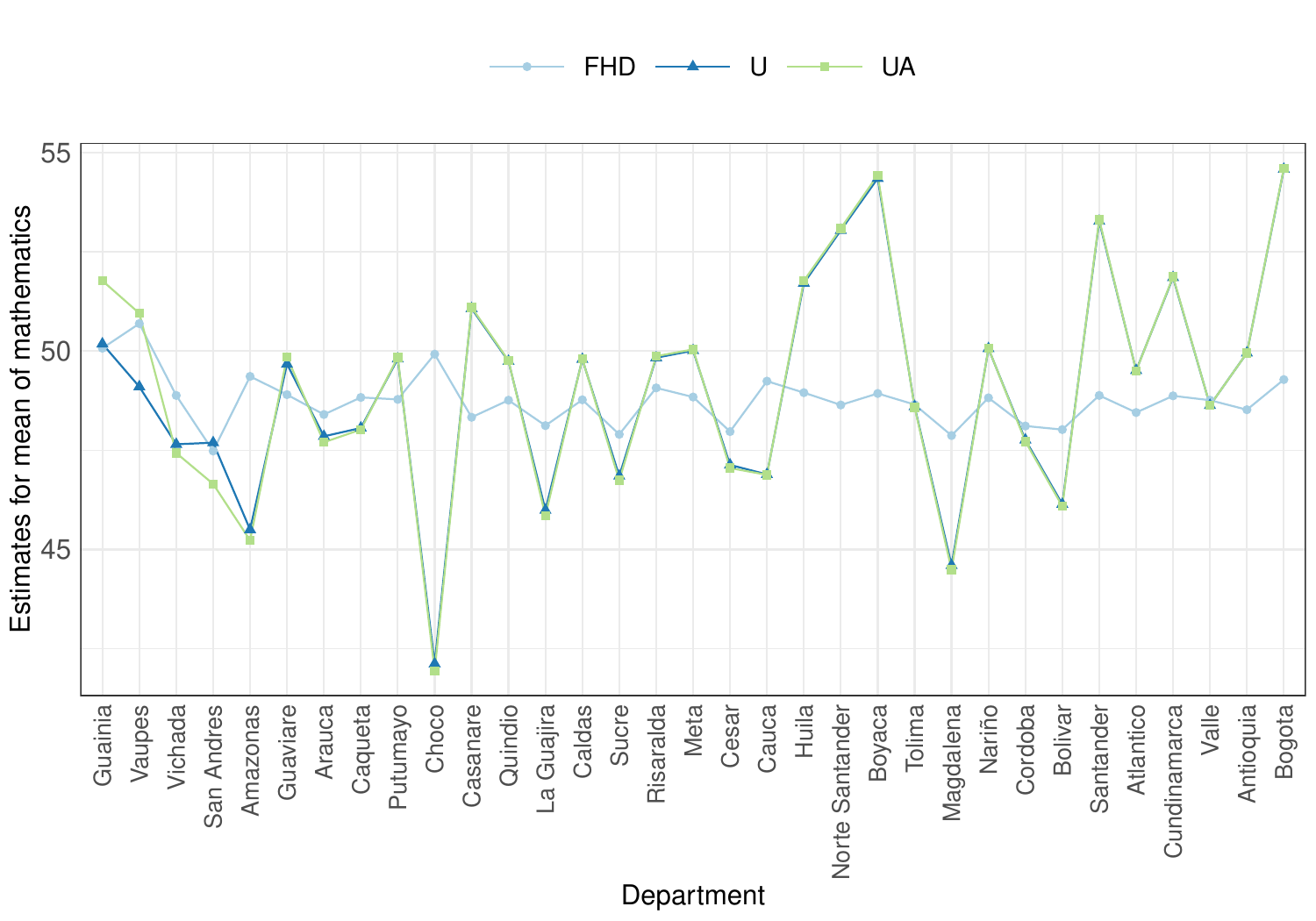}
	\caption{FHD, UA and U estimators of the means of Mathematics $\mu_d$ by department, in ascending order of sample size.}\label{fig:aplication_estimates}
\end{figure}

Given that no serious departures from the unit-level BHF model are found, we depict in Figure \ref{fig:aplication_estimates} the values of the model-based estimators $\hat{\mu}_{d}^{FHD}$, $\hat{\mu}_{d}^{UA}$ and $\hat{\mu}_{d}^{U}$ of the Maths department means, $\mu_d$, with departments sorted in the ascending order of sample size. In this application, $\hat{\mu}_{d}^{UA}$ and $\hat{\mu}_{d}^{U}$ take practically the same values, which differ from $\hat{\mu}_{d}^{FHD}$.

\begin{figure}[!h]%
	\centering
	\includegraphics[width=155mm]{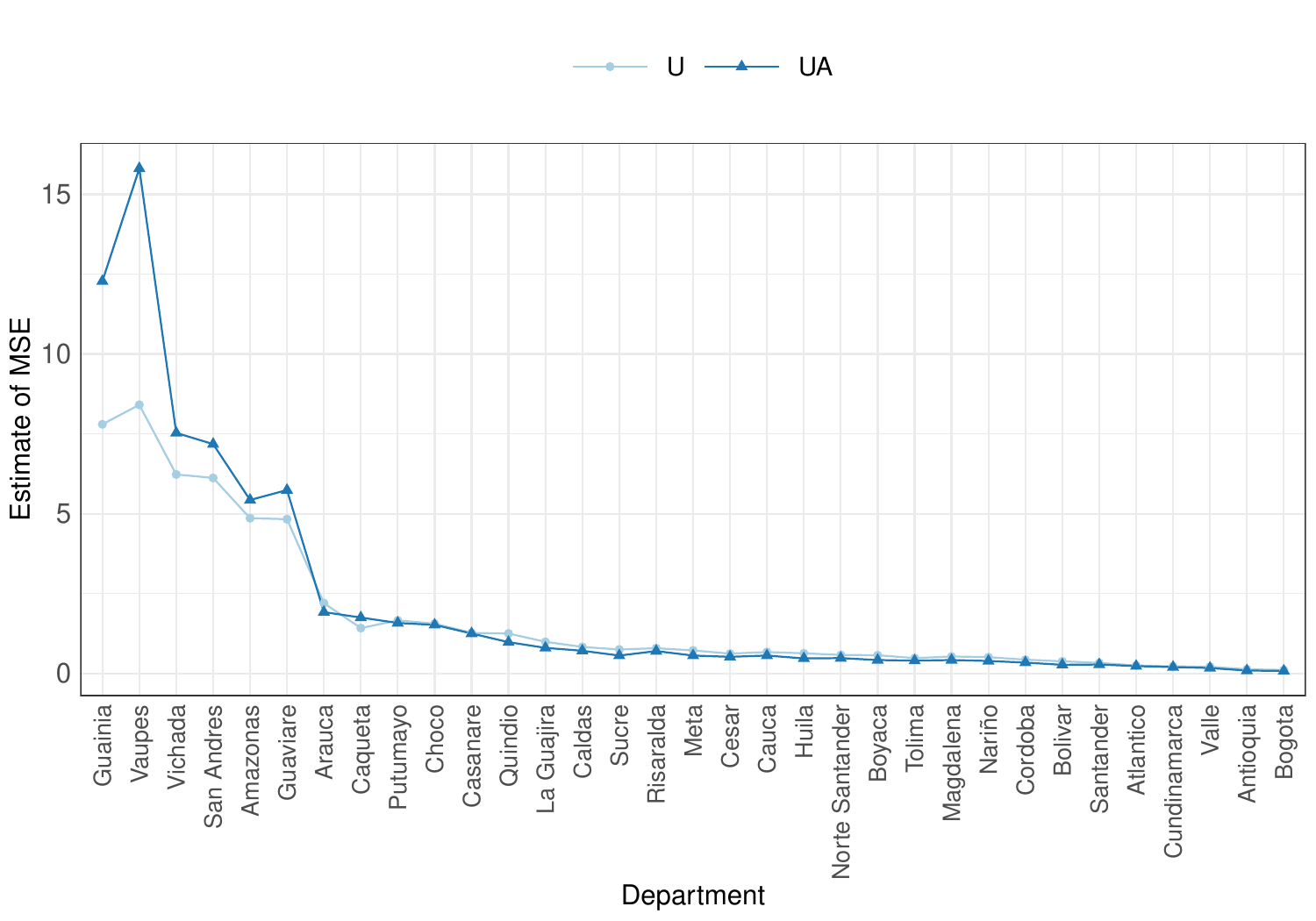}
	\caption{MSE estimates for UA and U estimators of the mean Maths scores $\mu_d$ by department, in the ascending order of sample size.}\label{fig:mse_aplication}
\end{figure}

Table \ref{tab:Tab2} in the Appendix reports summaries of MSE estimates for FHD, UA and U. We can see the larger MSE estimates obtained by FHD compared with UA and U. The maximum estimated MSE for U is almost half of that for UA, and that of UA is also clearly smaller than for FHD.
Estimated MSEs of the two EU estimators are shown for each department in Figure \ref{fig:mse_aplication}, in the ascending order of sample size. FHD is not displayed here because its much greater MSE estimates distort the plot. We can see that the estimated MSEs of UA are larger than those of U for the departments with the smaller samples sizes, but they get close as the sample size increases.

\section{Conclusions}\label{sec6}

This paper unifies the small area estimators of area means $\mu_d$ based on the most common unit-level and area-level models, by previously performing calibration of the survey weights to the area counts and totals of the auxiliary variables. The unified model can be fitted with data at the area or at the unit levels, requiring exactly the same data sources as the area-level FH model. The unified model allows us to estimate consistently the error variances in the FH model, and this in turn allows us to construct consistent MSE estimators that take into account the uncertainty due to the estimation of these error variances.

Our simulation experiments show that the empirical unified predictors  obtained with consistently estimated error variances, either using area-level or unit-level data, $\hat\mu_d^{UA}$ and $\hat\mu_d^{U}$, perform clearly better, in terms of RB and RRMSE, than the usual FH estimator $\hat\mu_d^{FHD}$ based on direct estimators of the error variances. Our simulations illustrate that $\hat\mu_d^{FHD}$ might be even less efficient than the direct estimator $\hat\mu_d^{DIR}$ for some areas.

Our bootstrap estimators of $\mbox{MSE}(\hat\mu_d^{U})$ perform clearly better than the analytical MSE estimator that ignores the uncertainty due to the estimation of the error variances. The latter overestimates the true MSE, specially in areas with small sample sizes. In contrast, our MSE estimators track well the corresponding true MSEs of the empirical unified predictors $\hat\mu_d^{UA}$ and $\hat\mu_d^{U}$ for all the areas, including those with very small sample sizes.
\bibliographystyle{agsm}

\bibliography{reference}

\newpage

\appendix

\section{Proof of design consistency of $\hat\mu^{U}_d$}

We proof that, if $w_{di}^C>0$ for all $i\in s_d$ and $\max_{i\in s_d}(w_{di}^C/N_d)=O(n_d^{-1})$, then $\gamma_d^C(\sigma_u^2,\sigma_e^2)\to 1$ as $n_d\to\infty$.
We have that
\begin{align*}
\sum\limits_{i \in s_d} \left( \frac{w_{di}^C}{N_d} \right)^2 &\leq n_d \max\limits_{i \in s_d} \left( \frac{w_{di}^C}{N_d} \right)^2 \\
&= n_d \left\{ \max\limits_{i \in s_d}  \left(\frac{w_{di}^C}{N_d} \right) \right\}^2 \\
&= n_d O(n_d^{-2}) \\
& = O(n_d^{-1}).
\end{align*}
As a consequence, $\psi_d^C(\sigma_e^2)=\sigma_{e}^2\sum\limits_{i \in s_d} (w_{di}^C/N_d)^2 \to 0$ as $n_d\rightarrow \infty$, since $\sigma_{e}^2 \in (0, \infty)$ is fixed. Therefore, $\gamma_d^C(\sigma_u^2,\sigma_e^2)=\sigma_{u}^2/(\sigma_{u}^2 + \psi_d^C(\sigma_e^2))\to 1$ as $n_d\rightarrow \infty$, since $\sigma_{u}^2 \in (0, \infty)$ is also fixed.

\section{Proof of selfbenchmarking property of $\hat\mu^{U}_d$}

For given $\sigma_{u}^{2}$ and $\sigma_{e}^{2}$,  $\tilde{\boldsymbol{\beta}}_U^C(\sigma_{u}^{2},\sigma_{e}^{2})$ is the solution of the following equation system for $\bbeta$:
\begin{equation}
\sum_{d=1}^{D}\sum_{i \in s_d} w_{di}^C\boldsymbol{x}_{di}[y_{di}-\boldsymbol{x}_{di}'\boldsymbol{\beta} - \tilde{u}_{dw}^C(\boldsymbol{\beta},\sigma_{u}^{2},\sigma_{e}^{2}) ]=\cero_p,
\label{BetaEquation}
\end{equation}
where $\tilde{u}_{dw}^{C}(\boldsymbol{\beta}, \sigma_{u}^{2}, \sigma_{e}^{2})=\gamma_d^C(\sigma_u^2,\sigma_e^2)(\bar{y}_{dw}^{C}- (\bar{\bx}_{dw}^{C})' \boldsymbol{\beta})$ and $\boldsymbol{x}_{di}=(x_{di1},\ldots,x_{dip})'$ with $x_{di1}=1$.
Taking the first equation in \eqref{BetaEquation}, corresponding to $x_{di1}=1$, and noting that, by the calibration restrictions, $\sum_{d=1}^{D}\sum_{i \in s_d} w_{di}^{C}\boldsymbol{x}_{di} = \sum_{d=1}^{D} \boldsymbol{X}_d = \boldsymbol{X}$, it holds
\begin{align*}
    \sum_{d=1}^{D}\sum_{i \in s_d} w_{di}^{C}y_{di} - \sum_{d=1}^{D}\sum_{i \in s_d} w_{di}^{C}\boldsymbol{x}_{di}'\tilde{\boldsymbol{\beta}}_U^{C}(\sigma_{u}^{2},\sigma_{e}^{2}) &= \sum_{d=1}^{D}\sum_{i \in s_d} w_{di}^{C} \tilde{u}_{dw}^{C}(\tilde{\boldsymbol{\beta}}_U^{C}(\sigma_{u}^{2},\sigma_{e}^{2}),\sigma_{u}^{2}, \sigma_{e}^{2}) \\
    \Leftrightarrow\hat{Y}_{w}^{C} - \boldsymbol{X}'\tilde{\boldsymbol{\beta}}_U^{C}(\sigma_{u}^{2},\sigma_{e}^{2})&=  \sum_{d=1}^{D} N_d \tilde{u}_{dw}^{C}(\tilde{\boldsymbol{\beta}}_U^{C}(\sigma_{u}^{2},\sigma_{e}^{2}),\sigma_{u}^{2},\sigma_{e}^{2}).
\end{align*}
 This is also valid for $\hat{\sigma}_{u,U}^{2}$ and $\hat{\sigma}_{e,U}^{2}$, and denoting $\hat\gamma_d^C=\gamma_d^C(\hat\sigma_{u,U}^{2},\hat\sigma_{e,U}^{2})$, $\hat{\boldsymbol{\beta}}_U^{C}=\tilde{\boldsymbol{\beta}}_U^{C}(\hat\sigma_{u,U}^{2},\hat\sigma_{e,U}^{2})$ satisfies
\begin{equation*}
\hat{Y}_{w}^{C} - \boldsymbol{X}'\hat{\boldsymbol{\beta}}_U^{C} =  \sum_{d=1}^{D} N_d \hat{u}_{dw}^{C},
\end{equation*}
where $\hat{u}_{dw}^{C} = \tilde{u}_{dw}^C(\hat{\boldsymbol{\beta}}_U^{C},\hat{\sigma}_{u,U}^{2}, \hat{\sigma}_{e,U}^{2})$. Then, we have
\begin{align*}
 \sum_{d=1}^{D} N_d \hat\mu_{d}^{U} &= \sum_{d=1}^{N_d} N_d\left[\bar{\boldsymbol{X}}_d'\hat{\boldsymbol{\beta}}_U^{C} +  \hat\gamma_d^{C}(\bar{y}_{dw}^C - \boldsymbol{X}_d'\hat{\boldsymbol{\beta}}_U^{C}) \right] \\
 &= \sum_{d=1}^{D}N_d \bar{\boldsymbol{X}}_d'\hat{\boldsymbol{\beta}}^{C}_{w} + \sum_{d=1}^{D}N_d \hat{u}_{dw}^{C} \\
 &= \boldsymbol{X}'\hat{\boldsymbol{\beta}}_U^{C} + \hat{Y}_{w}^{C} - \boldsymbol{X}'\hat{\boldsymbol{\beta}}_U^{C} \\
 &= \hat{Y}_{w}^{C}.
\end{align*}

\newpage

\section{Additional simulation results}

\begin{figure}[ht]%
	\centering
	\includegraphics[width=160mm]{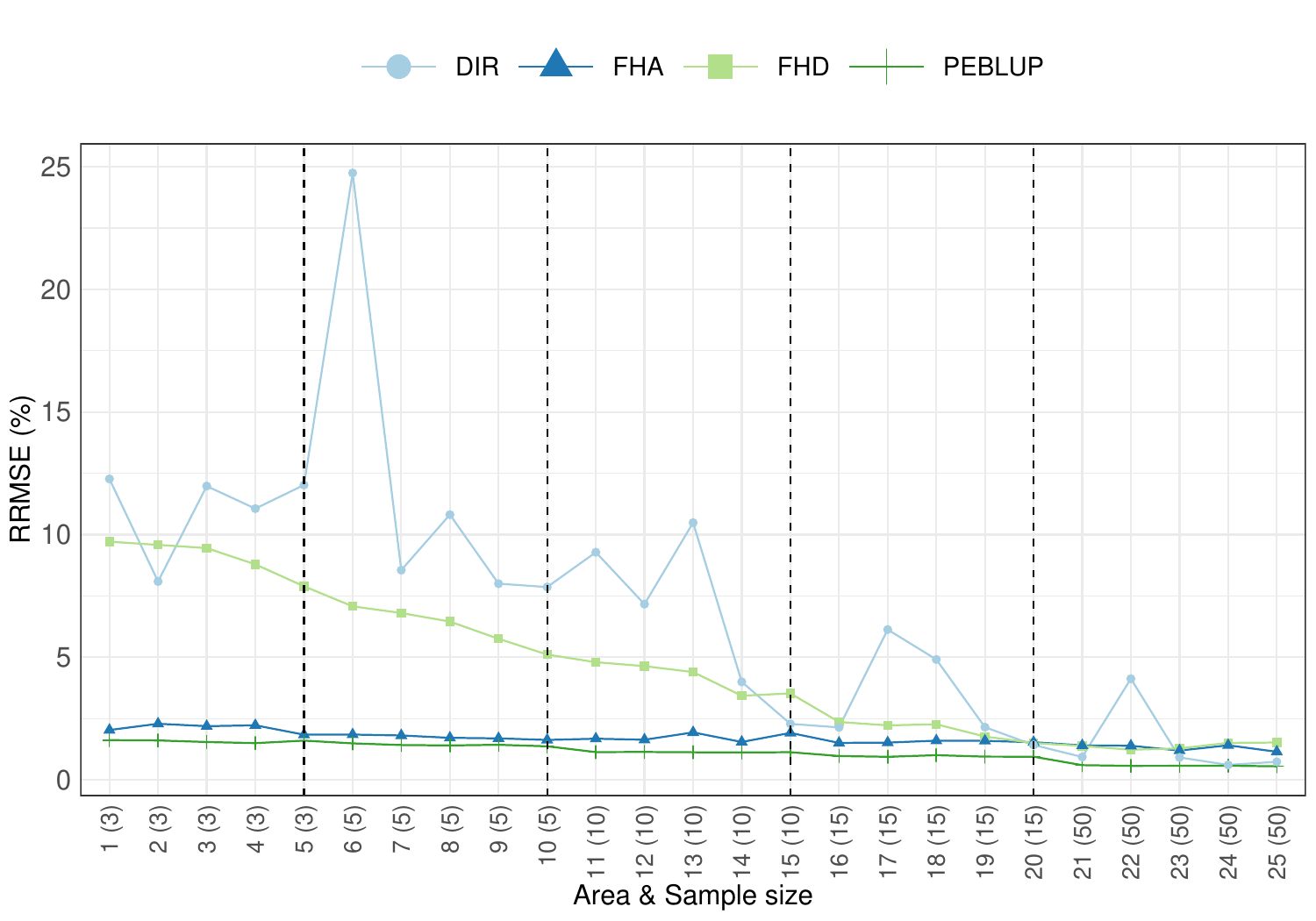}
	\caption{\% RRMSE of DIR, FHA, FHD and U estimators with original survey weights, for each area, with area sample sizes in parentheses.}\label{fig:mseFHTplot_SRSWOR}
\end{figure}

\newpage

\section{Additional application results}

\begin{figure}[!h]%
	\centering
	\includegraphics[width=150mm]{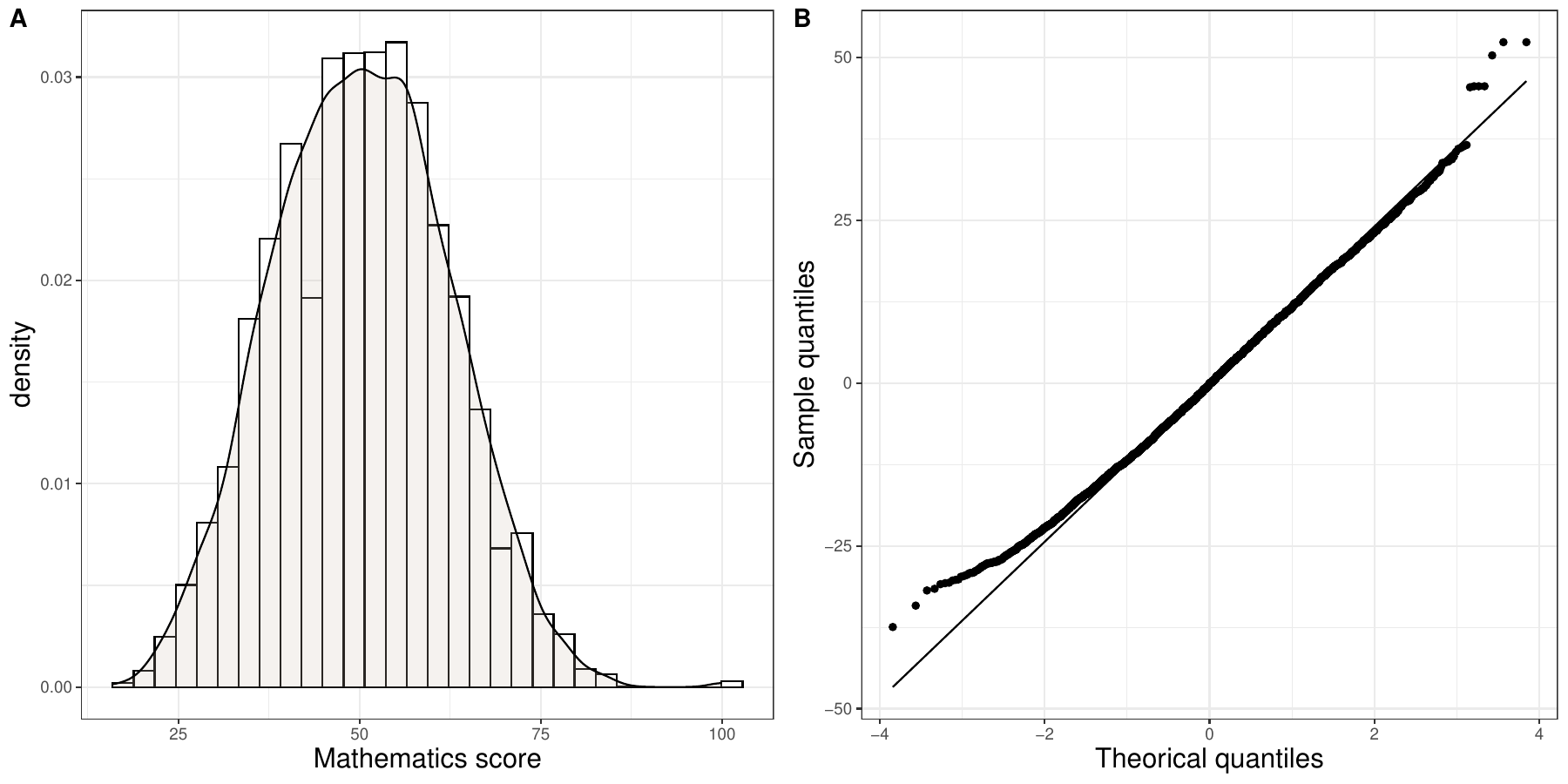}
	\caption{Histogram (A) and normal Q-Q plot (B) of unit-level residuals from BHF model.}\label{fig:residuals_model}
\end{figure}

\begin{table}[!h]
	\caption{Summary of MSE estimates for FHD, UA and U estimators of Maths mean score.}
	\label{tab:Tab2}
	\centering
	\begin{tabular}{lccccccc}
		\toprule
		Estimator & Min. & Q1 & Median & Mean  & Q3 & Max. \\ \midrule
		$\mbox{mse}_{PR}(\hat{\mu}_{d}^{FHD})$  & 6.83 & 7.23 & 7.58 & 8.79 & 9.57 & 20.60 \\
		$\mbox{mse}_{PB1}(\hat{\mu}_{d}^{UA})$ & 0.08 & 0.39 & 0.56 & 2.17 & 1.58 & 15.80 \\
		$\mbox{mse}_{PB}(\hat{\mu}_{d}^{U})$   & 0.11 & 0.48 & 0.72 & 1.77 & 1.56 & 8.41 \\
		\bottomrule
	\end{tabular}
\end{table}

\end{document}